# Exploring the visualisation of hierarchical cybersecurity data within the Metaverse

Terence Eden

2022-12

https://orcid.org/0000-0002-9265-9069

Supervisor Name: Dr Sharon Moyo

# Acknowledgements

This dissertation would not have been possible without the kind and generous support of several people.

My sincere gratitude to my academic supervisor Dr Sharon Moyo for her knowledge, encouragement, and patience.

Throughout this MSc programme I have been supported by my employer, the Cabinet Office. In particular, I would like to thank my line managers Arnau Siches and Simon Foster for giving me the space to study while working. A special mention to all my colleagues in Securing Government Services for volunteering to be guinea-pigs in my experiments.

Technical advice was magnanimously provided by Kaur Kullman from University of Maryland, whose work on Multidimensional Data Visualisations for Cybersecurity was a constant source of inspiration. Similarly, thanks to Peter Daukintis and the Mixed Reality team from Microsoft who graciously lent their time and expertise to discuss Metaverse hardware. Thanks also to Brian Eppert and the team behind Noda for building such an easy to use VR toolkit.

This project is built using a number of Open Source tools. My heartfelt appreciation to the authors of and contributors to D3.js, git, GNU/Linux, NumPy, Python, and Zotero.

Finally, eternal thanks to my wife Elizabeth Eden MA for her support during this MSc.

# Executive Summary

The author's employer is a part of the UK Government's Cabinet Office. They have a large amount of cybersecurity data relating to the Domain Name System. This project assesses the feasibility of viewing and manipulating the data in a 3D Virtual Reality environment known as the Metaverse.

This paper is the capstone project for an MSc Digital Technology Specialist qualification. As a project, it is designed to showcase a substantial piece of independent work which solves a real-world problem for the author's employer. It provides evidence of the author's professional development and demonstrates their expertise in the field of data analytics.

The primary aim was to extract cybersecurity data relating to domain names, create a 3D visualisation of the data, and build a prototype Metaverse environment in which to test an interactive model with real users. A further aim was to solicit participants' feedback and use this to assess what problems may be faced by organisations moving their workforce into the Metaverse. These aims were successfully met.

A prototype Metaverse experience was created in which users could explore hierarchical cybersecurity data. A small group of participants were surveyed on their attitudes to the Metaverse. They then completed a short series of tasks in the environment. Questions were asked to assess if they were suffering from Cybersickness. After completing further tasks, their attitudes were surveyed regarding future uses of the metaverse in the organisation. A second cohort of participants attended an online seminar. They completed a survey about their attitudes to the Metaverse. They then watched a short video of the Metaverse experience. Afterwards, they answered questions related to their attitudes towards future uses of the metaverse in the organisation. The results of these questionnaires were assessed to see whether participants were receptive to the idea of working with data inside the Metaverse in the future.

Of the 8 participants who tried the VR experience, the majority had little to no experience with VR until this study. All were able to complete data navigation and interpretation tasks in the Metaverse. The majority expressed excitement and interest in using VR in the workplace. No participant suffered from Cybersickness.

The 45 participants who watched the video demonstration of the Metaverse experience were more likely to have had previous experience with VR. They expressed slightly less interest in the way the Metaverse could be used in their organisations and also had concerns about the company who is developing the Metaverse.

In conclusion, while participants showed a clear interest in using VR to manipulate data, further work is needed to make sure future Metaverse experiences are accessible and do not endanger the health of participants. Several participants expressed distrust about Meta - the company formerly known as Facebook - which develops the Metaverse. The Metaverse is still in its infancy and more research needs to be done to find ways to make it more accessible and acceptable for users.

This MSc was funded and approved by the author's employer in line with the UK Government's Civil Service Apprenticeship Strategy ([Lopez and Chisholm, 2021](#)).

# List of Figures

# List of Tables

# List of Acronyms

- CSV - Comma Separated Values. A common data transfer format.
- Cybersickness. Ill health reaction caused by exposure to Virtual Reality.
- DAG - Directed Acyclic Graph. A data structure which defines relations between data points (known as nodes).
- DNS - the Domain Name System. The set of protocols which allows Internet Protocol addresses to be translated into memorable text strings.
- FPS - Frames Per Second. How many times a display is refreshed with a new image.
- HMD - Head Mounted Device. The generic term for a virtual reality headset.
- IPD - InterPupillary Distance. How close a human's eyes are to each other.
- PoC - Proof of Concept. A demonstration project which assesses the viability of an idea.
- VR - Virtual Reality. An immersive 3-dimensional environment facilitated by stereoscopic displays.
- VRSQ - Virtual Reality Sickness Questionnaire. A set of questions designed to assess whether a user is suffering from VR-induced Cybersickness.
- WebVR - Virtual Reality delivered via web technology. A new standard to allow VR experiences to be created and used with common web technologies.

# 1. Introduction

## 1.1 Context

The author works for a department of the UK Government which has a large amount of cybersecurity data relating to the Internet's Domain Name System (DNS). The data are hierarchical, complex, and classified as Secret. Because of this complexity and volume, there is a pressing need to make it easier for employees to be able to easily view and manipulate data in order to understand the issues represented.

This project aims to research ways to visualise complex hierarchical data in the virtual and interactive three-dimensional environment known as the "Metaverse".

The use of consumer-level gaming technology to create interactive 3D data representations is not new. The L3DGEWorld project used the "Quake III" gaming engine to power a real-time and collaborative environment ([Harrop and Armitage, 2006](#)), albeit one which could only be experienced via a traditional computer monitor. This project aims to extend that research into "Virtual Reality" (VR) by having participants wear a stereoscopic Head Mounted Display (HMD).

The UK Government has a long history of storing, analysing, and visualising complex data. In 1086CE, the Domesday book collated data about the nascent state of England and Wales. Because the data were recorded in textual form, there was no way to visualise or explore the data ([Hamshere and Blakemore, 1976](#)). In 1858CE, Florence Nightingale revolutionised the way the state's statistics were visualised through her use of two-dimensional Polar Area Diagrams ([Anderson, 2011](#)). A further paradigm shift is now underway regarding the display and interpretation of statistics about the state's infrastructure.

## 1.2 Potential Benefits

Cybersecurity data is crucial to the running of the author's department. They have a mission to safeguard the Government's presence on the Internet. This research assesses whether the data can be represented in 3D and how employees feel about using the Metaverse to navigate such data. By its nature, DNS information is hierarchical (see [Figure 06 - Illustration of the DNS hierarchy](#)) and so this research is limited to data where every element has a well defined "ancestor". Hierarchical data lends itself well to being presented in an interactive 3D environment where specific elements can be collapsed and hidden. Consequently, this research could be applied to any hierarchical data set, not just cybersecurity related.

Engaging with a new technology such as the Metaverse is risky for any organisation. As well as the costs of procuring hardware and developing software, there is the risk that employees will find the new way of working to be uninteresting or unhelpful. There is also a risk that VR may cause employee sickness. The benefits of this research to the author's organisation will be a better understanding of these risks.

## 1.3 Main Aims

This work contains three main aims:

1. To build a prototype which allows users to visualise complex hierarchical data in the Metaverse.
2. To determine users' attitudes to using the Metaverse in a professional context, both before and after experiencing the prototype.
3. To investigate whether "Cybersickness" occurs with users who try the prototype.

Usable 3D computer interfaces have been a highly anticipated development to the point where Hollywood's vision of how future computers will act has become a meme ([Grazier and Cass, 2017](#)) as can be seen in [Figure 01](#). But are such systems practical? Do workers feel excited or concerned about using these interfaces? Are there Health & Safety issues which need to be addressed?

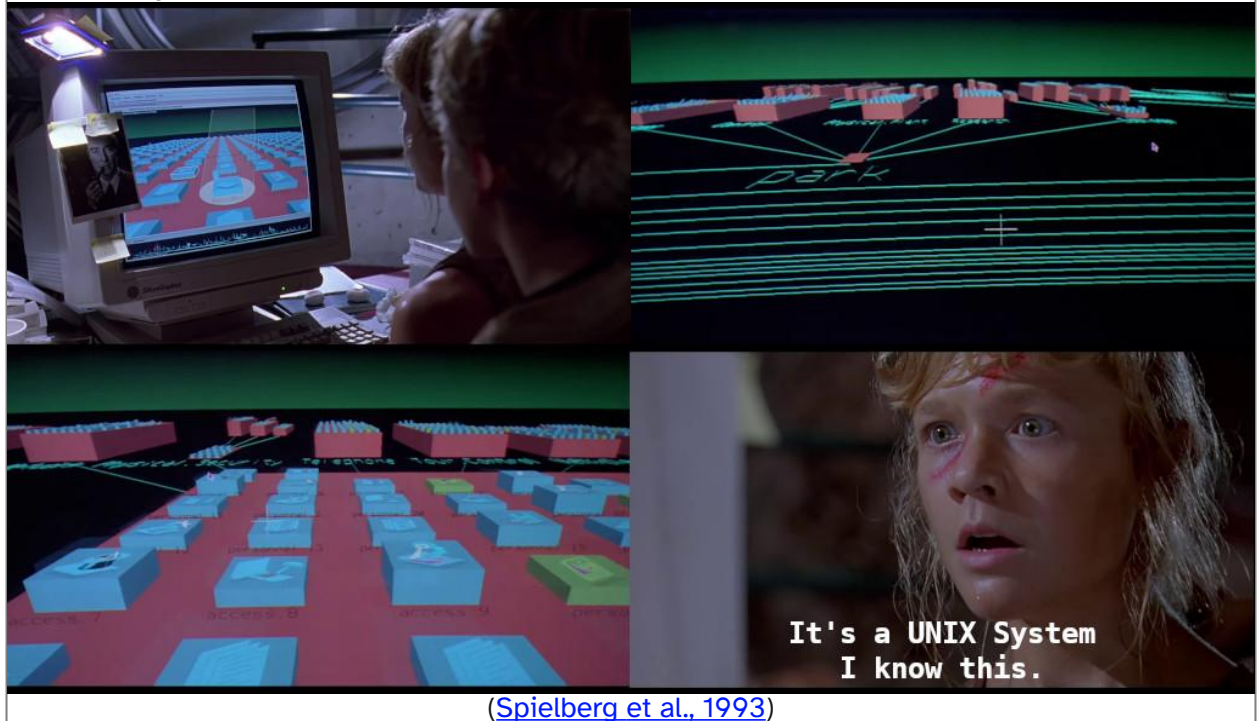
Figure 01 - A fictional depiction of 3D visualisations of Cybersecurity interfaces
(Spielberg et al., 1993)

These questions inform the project's objectives.

## 1.4 Objectives

The original objectives from the project's Terms of Reference were to:

1. Investigate current literature and identify modern techniques for 3D interactive visualisations.
2. Gather data related to the domain name hierarchy and its security issues.
3. Experiment with data structures which can be easily visualised in 3D to allow others to analyse the data.
4. Report on whether the visualisations make for more engaging and useful presentations for stakeholders.

Following the literature review and discussion with stakeholders, a further objective was added: 5. Discover whether people suffered any ill-health effects while using the visualisations.

## 1.5 Professional Issues

The author's organisation deals with both secret and top secret data. There is a drive to make these data more open and to change the way such data are classified (Heide, 2022). Until such changes can be made, the data must be kept secure at all times. This placed limits on the machines that could be used to analyse the data. As such, data were encrypted both at rest and in transit between devices.

The organisation also has a well-developed "National AI Strategy" which emphasises the need to develop cyber-physical infrastructure (GOV.UK, 2021). This work has to support the broader strategy, and does so by performing in-person testing of receptiveness to the Metaverse among the workforce.

The author is a Member of the BCS and follows their code of conduct (British Computer Society, 2021). This was particularly relevant due to the code's insistence that members "have due regard for public health, privacy, security and wellbeing of others", which was the basis of the ethical approach used in designing the study. Similarly, the work was carried out under the Civil Service Code which insists on a rigorous approach to impartiality and "fiduciary obligations" (Fuertes, 2021). These codes informed the author's approach to the following issues.

## 1.6 Social, Ethical, and Legal issues

There are several Health & Safety issues associated with the use of HMDs. While wearing the device, users are oblivious to their surroundings, risking unexpected kinetic interactions with their environment. Therefore testing took place in a private room with adequate space for the participant to move around. To guard against other issues, a full Health & Safety risk assessment was conducted and is presented in [Annex 1 - Risk Assessment](#).

In the paper "Exploring the Unprecedented Privacy Risks of the Metaverse" ([Nair, Garrido and Song, 2022](#)), the authors present several ways in which VR can be used to infer personal data about users. Although the paper is a pre-print, and was only tested on a limited and non-diverse set of participants, it is this author's opinion that the paper identifies realistic threats. These must be taken into account by researchers in order to protect the personal data of their research subjects. As such, the prototype developed by the author does not collect any personal data.

Finally, ethical clearance for testing the prototype was granted by the University of Northumbria. All interview participants were presented with an information sheet about the aims and risks of the research, and gave verbal consent to participation & transcription of their interview.

## 1.7 Report Structure

This report begins with a literature review to examine current research on the Metaverse. It is followed by a discussion of the methodology used for the author's research and development. Next is a section on the practical research work undertaken including development of a workable prototype. The results from this practical work are then analysed and, in a further section, critically evaluated. Finally, the author makes their conclusions and recommendations for further work.

# 2. Literature Review

## 2.1 Background

In order to survey the current state-of-the-art, the author performed a literature review. This review surveys a wide range of modern and historical papers relating to the project, and discusses their relevance, applicability, and utility. The review starts by examining foundational papers on the nature of the Metaverse. It then considers whether businesses find practical utility from Virtual Reality. Next, it discusses various data visualisation studies. It concludes with a synthesis of this knowledge and how that knowledge will be applied to this and future projects.

## 2.2 Scope

The literature review was conducted in line with modern expectations of both academia and industry (Berdanier and Lenart, 2021).

The Metaverse is an all-encompassing Virtual Reality (VR) environment which allows a user to interact with objects, data, and other users. The etymology comes from the Greek word "meta" ("beyond") and "verse" (short for "universe") meaning a reality beyond the physical plane of existence (Dionisio, III and Gilbert, 2013).

There is a growing field of study around the Metaverse and, specifically, how it relates to education and business processes. However, because this is a fashionable new field, there are many papers which use "Metaverse" as a keyword to increase their visibility in search engines, but barely relate to the topic. These papers have been excluded from this review. There are multiple papers looking at Metaverse from a gaming perspective, but these have only been included in this review where they also contain significant content about data structure visualisation.

There is a small but growing body of academic work looking at whether VR is a suitable environment in which to conduct business. Many papers are little more than paid-for editorials, often known as advertorials (Stirling, 2018). These papers prioritise their sponsor's viewpoint rather than an objective academic one. Due to this lack of academic rigour, these papers have been avoided where possible.

Much of the literature around data visualisation refers to either medical or financial data. Although not directly relevant to the visualisation of cybersecurity data, they are still of interest to see how professionals interact with 3D data.

## 2.3 Metaverse

The term "Metaverse" first appears in Neal Stephenson's influential 1992 cyberpunk novel "Snow Crash" (Zyda, 2022). This term quickly found its way into academic literature discussing the possibilities of creating interactive 3D worlds (Parr and Rohaly, 1995). These early papers were mostly focussed on the mechanics of creating such environments and the syntax used to describe them. Although technically detailed, they fail to look at concerns like usability, accessibility, and the ethical implication of their designs.

Since the Facebook corporation rebranded itself as "Meta" in 2021 (Egliston and Carter, 2021), the hype around the term "Metaverse" has been disproportionate to the reality of its impact (Dwivedi et al., 2022). The word Metaverse is now frequently misapplied in order to increase engagement (see Figure 02) with one commentator remarking that the term Metaverse was so over-used and over-hyped that it was treated as "a punchline" (Ball, 2022).

Figure 02 - Technology terms used in startup descriptions and tech articles

## Technology terms used in startup descriptions and tech articles

| 2020 | 2021 |
|---|---|
| Multiplayer game | Metaverse |
| Virtual Reality experience | Metaverse |
| Augmented Reality filter | Metaverse |
| 5G Connection | Metaverse |
| AR Cloud | Metaverse |
| Digital Avatar | Metaverse |
| Digital Event | Metaverse |
| ML classifier | Metaverse |
| E-commerce | Metaverse |
| Blockchain | Metaverse |
| Internet | Metaverse |
| Social Media | Metaverse |
| Videocall | Metaverse |
| Porn | Metaverse |
| Potato | Metaverse |

Humorous graphic (Vitillo, 2021).

Reviewing the relevant literature led the author to conclude that the Metaverse has become a catch-all term for a variety of VR-related technologies. Despite the efforts of Meta (née Facebook) to colonise the term, there is a thriving ecosystem of organisations which are using "Metaverse" as a generic term for modern, interoperable VR.

The author's working definition of the Metaverse follows this modern usage; a generic term for an interactive 3D environment experienced through Virtual Reality hardware. Conducting the literature review has highlighted the risk that the project will be inaccurately associated with Meta's commercial products.

## 2.4 Business in VR

There is a demand for businesses to be able to host meetings in virtual space (Pearlman and Gates, 2010). While their study did not perform any analysis of actual activity taking place in VR, the authors were able to identify the existence of VR as a viable route for business events through their qualitative research. However this viability is undermined by their findings showing a low level of penetration; only 2% of respondents actually used VR for their events. Another significant finding was the lack of interoperability between VR solutions. This is corroborated by a subsequent paper (Dionisio, III and Gilbert, 2013). As noted in Pearlman and Gates (2010), widespread adoption was limited by the primitive state of technology. Neither paper addresses how improved graphical fidelity and faster speeds could improve the experience.

In the field of business education Metaverse training may have many significant advantages over traditional asynchronous educational videos (Lee, Woo and Yu, 2022). It is important to note that their study's participants were recruited and tested via video-chat. The author contends that the selection of these participants would have been biassed towards those already familiar with virtual experiences, which may affect the applicability of their results. The paper concludes that interactive 3D content has both greater visual appeal and greater semantic content compared to data presented in tabular format. This validates the findings from earlier studies (Tanlamai,

Savetpanuvong and Kunarittipol, 2011). Many of these studies assert that the novelty of interactive content may have made participants more accepting of the experience. In this author's opinion this assertion is unsupported because, for many participants, interactive 3D objects are already an established part of their media landscape, for example in video games. Additionally, most papers highlight that the poor user experience of current technology may cause negative sentiment in participants without acknowledging that the user experience is likely to improve as technology advances.

Having reviewed the literature, the author concludes that it is important to calibrate for participants' pre-existing familiarity with VR in order to understand whether they value the novelty more than the utility of the final product. Therefore, the project will assess each participant's prior experience with the Metaverse and consider if this affects the results.

## 2.5 Data Visualisation

There is excitement in the field of medical research at being able to visualise the data structures of biological compounds (Taylor and Soneji, 2022). This is corroborated by studies on experienced surgeons, which found that the ability to visualise three-dimensional imaging resulted in fewer errors than conventional 2D visualisations (Timonen et al., 2021). Although statistically valid, the study on surgeons was only conducted on a small cohort (5 participants) who received a short period of VR training (15 minutes). A further weakness was that this study only looked at experienced surgeons. It did not investigate whether less experienced users would also benefit from this form of data visualisation.

A recent study of corporate annual reports shows how the design of charts in financial reporting can potentially mislead the reader (Shen, Lee and Wang, 2020). The study goes into detail about how the æsthetic choices can greatly influence the perception of the data, for example rendering graphs in an isomorphic fashion to simulate three-dimensionality. A later study corroborates that the design of graphs can mislead the viewer (Woodin, Winter and Padilla, 2022). However, the Woodin et al study was only concerned with 2D representation. The author contends that 3D images projected onto a 2D space are fundamentally different from structures which can be examined in their native projection. The review found there is a gap in the literature regarding the utility of true 3D visualisation experienced with stereoscopic equipment. This project addresses that gap.

This section of the literature review has had a positive impact on the project. The author is now aware that it is possible to create something visually appealing which nevertheless misleads users. The author is more cognisant of the need to develop a solution which faithfully renders the data.

## 2.6 Evaluation & Synthesis

In conclusion, multiple studies have shown the potential for VR to be a useful addition to the suite of tools used to visualise data. While the hype of the Metaverse shows no signs of diminishing, it is important to temper it with reality. Studies have shown that the novelty of VR often overshadows its utility. In addition many studies have failed to correct for the pre-existing expertise of test-subjects both in terms of their familiarity with the material and the VR environment.

One perennial problem is that authors seem wary of predicting how future improvements in technology will change the way users interact with the Metaverse. Moore's Law is now well established. Moore postulated that the power of computing doubles relative to price roughly every 18 months (Shalf, 2020). The author thinks it is reasonable to infer from this that reduced latency and increased bandwidth are inevitable. These improvements when coupled with smaller, lighter, and cheaper electronics should result in a better Metaverse experience.

Outside of the scope of this project, areas of future study should look at whether there is a significant disparity in the acceptability of VR between different age groups. It is also worth exploring whether people who have grown up with VR environments will be more willing to engage with the Metaverse in a professional context.

In conclusion, the author's research combines several aspects of the existing knowledge-base. This paper synthesises several modern research strands. By learning how other academics have approached these problems, this research project will investigate whether the Metaverse is an acceptable environment for exploring data-visualisation problems.

The following section describes the approach taken in this project.

# 3. Methodology

## 3.1 Introduction

This section is a brief critical review of the appropriate methodologies for both research and development. It discusses the author's research methods and plan for implementation.

## 3.2 Research Methods

The literature review was undertaken via a variety of Internet libraries including Google Scholar, IEEE Xplore, and Scopus. Keyword searches included VR, Metaverse, Augmented Reality, 3D statistics, Cybersickness, and their derivatives. Citation chains were followed to fully explore the relevance of the papers found. The author contacted leading experts in the field of Metaverse studies. Both academic and industry experts were interviewed to understand state-of-the-art research and discover upcoming publications. This approach helped the author find and read a wide range of papers.

## 3.3 Research Constraints

The design choice of a study is inexorably bound to both the aims of the research and the nature of the problems being investigated ([Walliman, 2010](#)). Because the solution being proposed is novel, many of the relevant works were pre-prints or PhD theses. Close investigation was needed to check the validity and replicability of their findings.

## 3.4 Analysis Methodology

Due to the fickle nature of quantitatively recording subjective topics such as attitudes, behaviours, and experiences, a study such as this one requires a qualitative approach ([Tenny et al., 2022](#)). Because of the novelty of this project, and the relatively small number of participants, any statistical analysis was likely to be inconclusive. Therefore, experiments and questionnaires were designed to capture participants' subjective feelings.

The study consisted of several elements. Firstly a research questionnaire was used to assess the participants' current knowledge of VR. Secondly, participants were asked to perform tasks in the Metaverse environment. While in the Metaverse, a Virtual Reality Sickness Questionnaire (VRSQ) was administered to ensure the safety and comfort of participants, and also to assess the risk of Cybersickness ([Kim et al., 2018](#)). Finally, a qualitative study explored their feedback on the VR experience. The research questionnaire was built around a five-point Likert Scale which is optimal for extracting the maximum reliability and validity from the answers ([Taherdoost, 2019](#)). It was important to design the survey to take into account how the experience of being asked questions alters the subjective experience of the respondent ([Scalia et al., 2022](#)).

One known limitation of asking post-experience questions is that it relies on participants' memories. This might provide an incomplete or inconsistent recollection of the experience ([Schwind et al., 2019](#)). Therefore, the study was designed to ask participants some questions while they were inside the Metaverse to capture their immediate thoughts, and then further questions once they had exited the experience.

Because qualitative research has a free-flowing structure and is exploratory rather than prescriptive, it provides insights into the scope of the problem and feeds into the future design of technology solutions ([Pearlman and Gates, 2010](#)). Due to the small sample size, the results were evaluated holistically rather than statistically. This sort of thematic analysis requires an analysis of the hierarchy of responses to interpret the data ([Castleberry and Nolen, 2018](#)).

Relying on participants' comments is sometimes seen as merely reflecting superficial opinions which are inferior to quantitative inquiry ([Morse, 2002](#)). However, the author disagrees with this assessment. In a small study, it is important to listen to participants' expressed feelings because they may not always be revealed by the limitations of a Likert Scale questionnaire. Research also finds that relying solely on reported quantitative results may not be appropriate for assessing visualisations ([South et al., 2022](#)).

## 3.5 Implementation methodology

Building the testable prototype followed the Agile methodology. A traditional "Waterfall" project follows a rigid development plan whereas Agile enables the project plan to adapt to any challenges which arise as the project proceeds. Woody Zuill, one of the founders of the Agile movement ([Gleeson, 2017](#)), is noted for his maxim "Doing exposes reality" ([Zuill, 2012](#)); the development of the prototype changed iteratively as relevant information was discovered during each Sprint (see [Table 01 - Initial Sprint Plan](#)). There are several criticisms of 2001's Agile "Manifesto"; modern researchers note that Agile prioritises the developers' experience at the expense of user experience, and Agile processes don't always fit modern business practices ([Ozkan, 2019](#)). However, this can be mitigated with the ScrumBan method which is recognised for improving quality and reducing wasted time ([Petricioli and Fertalj, 2022](#)) and is used in the author's workplace.

There were two cohorts of participants. The first were stakeholders from the author's immediate team who were asked to test the Metaverse prototype. Some were practitioners who work with DNS data as part of their roles and some were executives who need to understand, but rarely interact with, the data. The second cohort consisted of participants who chose to attend a virtual seminar. They saw a video demonstration of the Metaverse prototype. These participants were self-selecting members of the author's professional peer group. Each participant from both cohorts was given information about the project and its aims. They also provided their consent to participate in the study and have their answers recorded. Consent information can be found in [the appendix](#). Participants were not remunerated for their involvement.

Outside of the study, the work was presented to peers at a "Show and Tell". This is an established part of the Agile culture and a necessary step to obtain feedback from interested parties (Hulshult, 2019). Feedback from these presentations was also integrated into the prototype's design.

### 3.5.1 Sprint Plan

Agile projects are usually divided into shorter components called "Sprints". The project consisted of 6 Sprints, each around twenty-days.

| Table 01 - Initial Sprint Plan | |
|---|---|
| Sprint | Contents |
| 1. Academic Research (to feed in to literature review) | Prior Art - Interactive 3D visualisations of security related data<br>Usability - issues to be aware of in creating metaverse experiences<br>Psychology - how understandable are 3D graphs?<br>Ethical issues - cutting edge issues relating to the metaverse |
| 1. User Research | Identify Stakeholders<br>Gather user stories via interviews<br>Understand what risks the business wants to identify from the data |
| 1. Data Gathering | Get access to relevant systems<br>Extract data in a variety of forms<br>Transform data into different data structures<br>Load data into suitable systems |
| 1. Experiment with different visualisation techniques | Static / Non-interactive (i.e. data structure is rendered once)<br>Interactive (i.e. data structure can be manipulated)<br>Pre-recorded animation (i.e. a story is told with the data and visualised) |
| 1. Test with users | Qualitative questions<br>Feedback |
| 1. Report findings | Conclude write-up |
| 1. Submission | By 2022-12-07 due to travel commitments |

Sprints 2, 3, and 4 were planned to run simultaneously in order to produce a tightly coupled feedback loop.

As part of the Agile process, a Proof of Concept (PoC) was created to demonstrate the feasibility of the project. This was created in tandem with the Literature Review in order to establish how the project would likely proceed. This is described further in the next section.

# 4. Description of Practical Research Work Undertaken

This chapter describes the requirements analysis, system design, construction, and experimental work undertaken. It provides information about the user testing and data collection, and it justifies the techniques chosen. Finally, it addresses ethical and legal issues with the chosen approach.

## 4.1 Requirements Analysis

The author's workplace contains multiple stakeholders with an interest in DNS security. Given the wide range of potential stakeholders, it was useful to map them onto a Power Interest grid (see [Figure 03](#)) to understand their influence and engagement with the project ([Bernstein, Weiss and Curry, 2020](#)).

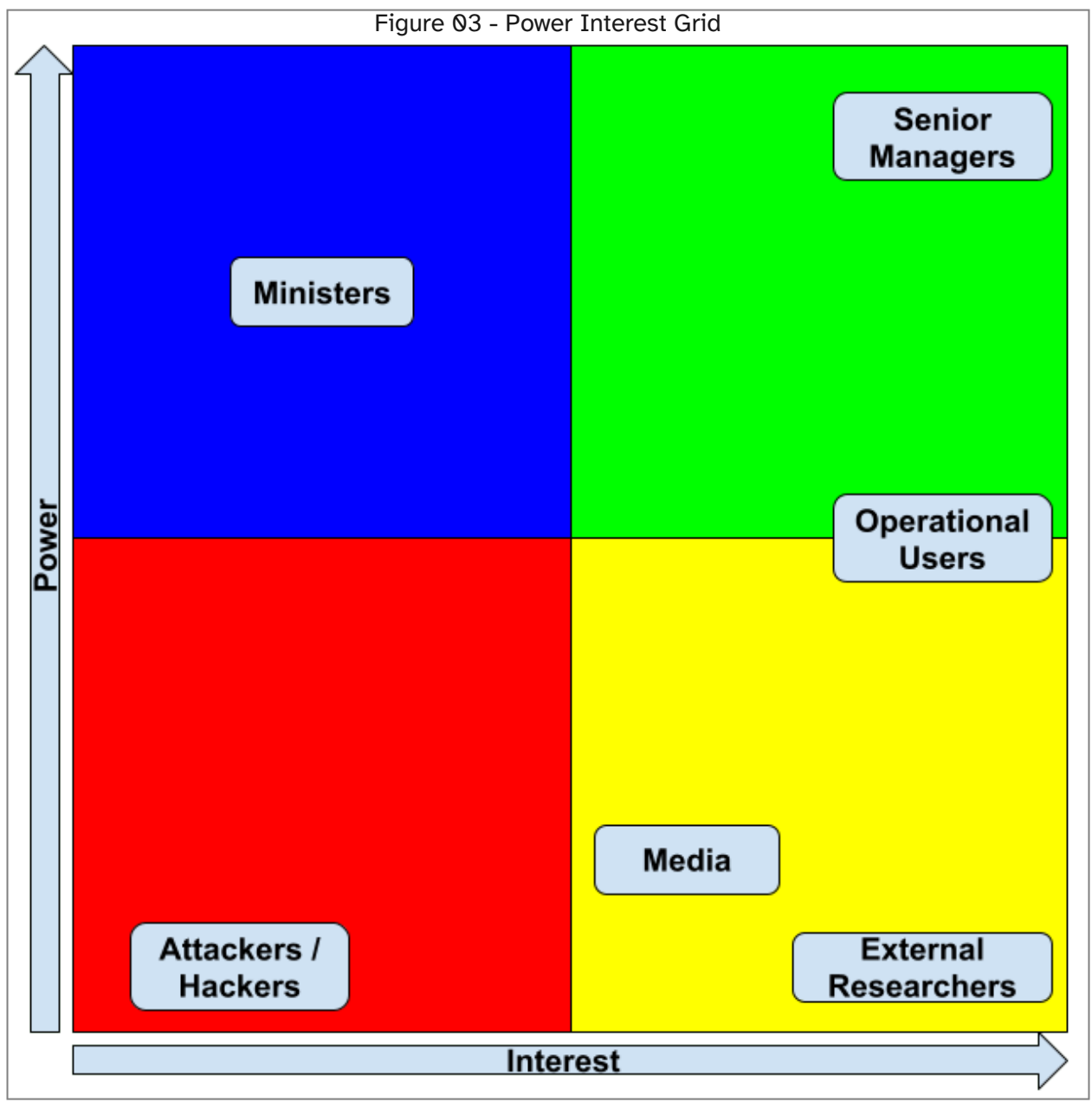

Figure 03 - Power Interest Grid

Although the two-dimensional nature of the grid limits the analysis to only two factors, Mendelow's grid makes it easy to see which actors need to be considered during the work. This form of analysis is accepted in the author's workplace to identify the target users. A discussion with the stakeholders revealed a pre-existing requirement to visualise the vast amount of data collected about the security of domain names, and see where in the hierarchy issues occur. The prototype was designed to fulfil this requirement.

## 4.2 Health and Safety

Any study which involves users requires a risk assessment to ensure their safety. Exposure to VR can cause disorientation, increased heart rate, and nausea. Collectively, these symptoms are referred to as "Cybersickness" and can cause distress in participants ([Petri et al., 2020](#)). Although Petri's study was small, the author has personally experienced Cybersickness and noticed it in others. There is a long history of research into human safety issues in VR, mostly around eye-strain and becoming entangled with trailing cables ([Viire, 1997](#)). While modern HMDs eschew external cables, there is still a risk of severe injury from unexpected encounters with the physical environment ([Warner and Teo, 2021](#)). The primary ethical concerns with this study were around Health and Safety, especially the risk from infectious diseases. Because the HMD and controllers were used by multiple participants, and given the HMD's proximity to a user's mucus producing orifices, enhanced cleaning of the equipment was necessary ([Moore et al., 2021](#)). A risk assessment was conducted in line with the author's employer's guidelines and is presented in [Annex 1 - Risk Assessment](#).

## 4.3 Ethics

An ethical review was conducted in line with the University's guidelines ([Northumbria University, 2022](#)). The ethical risks were judged low with one exception; the data used in the project are classified Government information with severe penalties for improper use ([‘Official Secrets Act’, 1989](#)). Consequently, participation was restricted to people with the relevant security clearance. It is a recognised problem that VR visualisations often require vast amounts of data, and those data are often transmitted to and stored in relatively insecure headsets ([Kohnke, 2020](#)). In order to avoid breaching the ethical guidelines around security, care was taken to protect the data in transmission and at rest - through encryption and cache minimisation respectively.

## 4.4 Design

The author collaborated with a number of external experts in order to validate that the technology existed to support this project. The Mixed Reality team at Microsoft helped the author understand what their Hololens product was capable of, but warned about the high cost of hardware and the rendering limitations of the platform. The author also interviewed Kaur Kullman, a noted expert in VR visualisations of cybersecurity data from University of Maryland. Kullman's research (see [Figure 04](#)) shows the viability of using interactive 3D environments to visualise complex cybersecurity information ([Kullman and Engel, 2022](#)).

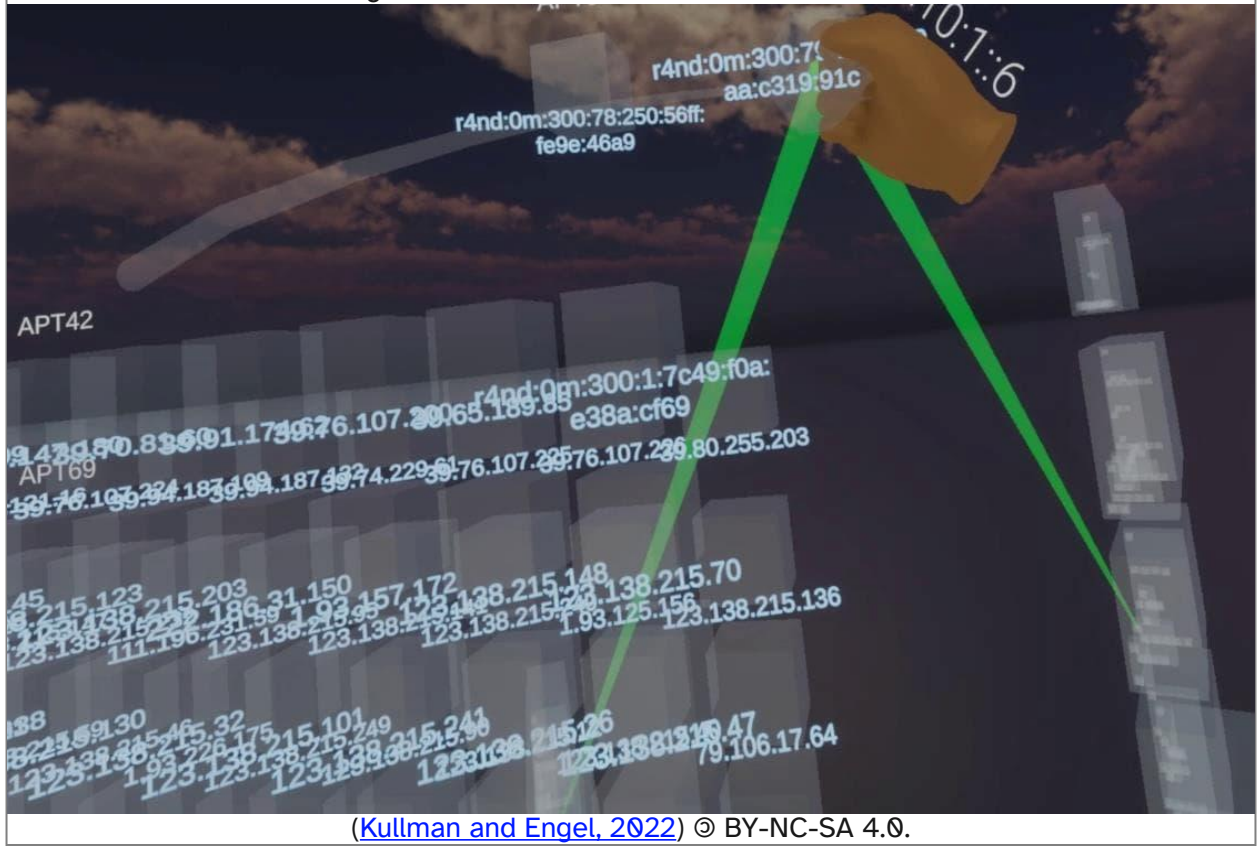

Figure 04 - Screenshot of 3D interactive data
([Kullman and Engel, 2022](#)) ⓒ BY-NC-SA 4.0.

Kullman explained the current state-of-the-art, potential hardware issues, and the physical limitations of human test subjects. This fed into the author's Sprint planning and project design. These collaborations informed the choice of hardware and platform.

Although there has been some academic analysis of consumer preferences for VR platforms ([Yang and Nam, 2018](#)), there is little research about enterprise needs. Therefore, the author undertook a conjoint analysis of the available hardware options.

|  | Table 02 - Conjoint Analysis of hardware choices | | | | | |
|---|---|---|---|---|---|---|
| Brand | Price | Performance | Weight | Programming Environment | Controls | Screen resolution (per eye) |
| Hololens (Microsoft, 2022) | High (£3,350) | High | 566g | Microsoft Windows Only | Hand tracking | 2048*1080 |
| Oculus Quest 2 (Meta, 2022) | Medium (£350) | Medium | 530g | Web, Android app, in-app tools | Hand tracking, joysticks | 1832*1920 |
| Cardboard VR (Google, 2022) | Low (£5 + cost of existing Android phone) | Low | <200g | Web, Android app | None | 1080*1080 |

The Oculus Quest 2 was chosen as the hardware platform as it offered the most flexible programming environment. It also had higher resolution, lower weight, and more control options than the alternatives. While the price was higher than the Cardboard, it was significantly lower than the Hololens.

## 4.5 Experimental Work

In order to build a performant model, several platforms were considered and tested. The first Sprint's aim was to build an experimental Proof of Concept (PoC) application. This PoC was created using a framework developed in conjunction with Mozilla called WebVR (Oliver et al., 2019). The code for the PoC is in Annex 2 - PoC Source Code. The DNS hierarchy is commonly described as a Directed Acyclic Graph (DAG) (Hoffman, Sullivan and Fujiwara, 2019), therefore, this representation was chosen for the visualisation. The PoC successfully loaded a small amount of DNS data from a CSV file, processed it into a DAG, and rendered it in interactive 3D (see Figure 05). This demonstrated that the project was conceptually viable.

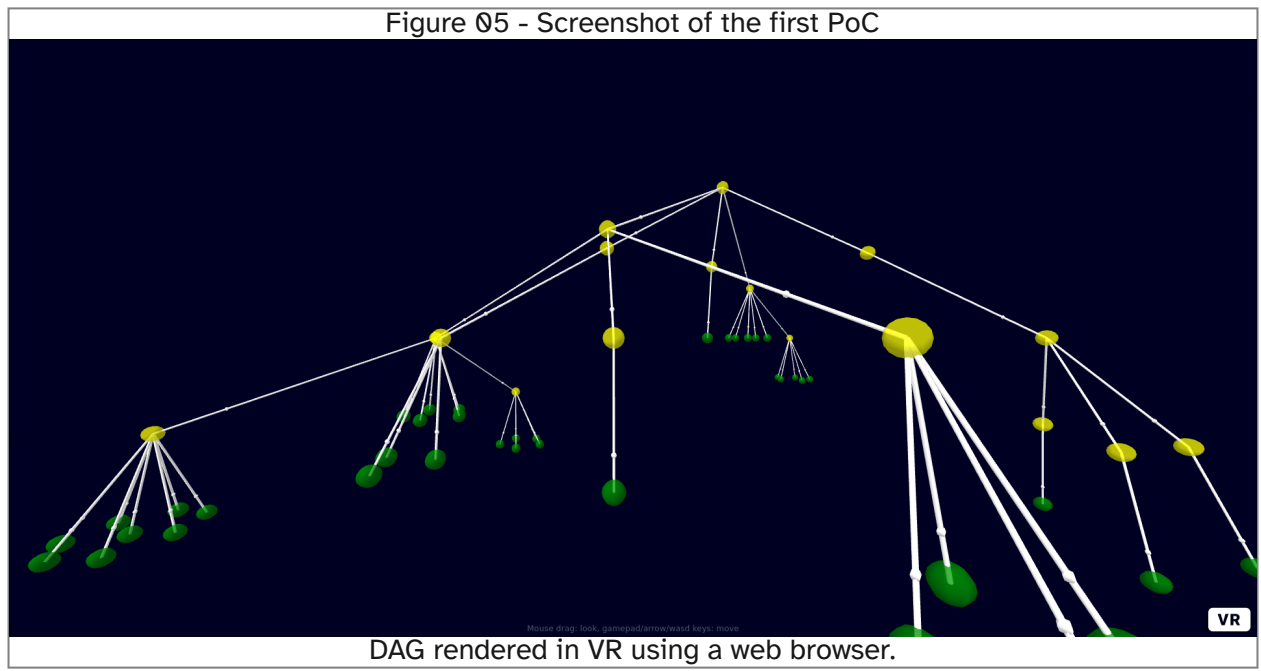

Figure 05 - Screenshot of the first PoC
DAG rendered in VR using a web browser.

The PoC was delivered to Microsoft who confirmed that, although it ran on their Hololens hardware, they noted that there were no native VR controls. An updated PoC was created to address this issue, and Microsoft reported that it ran satisfactorily. This meant that the next Sprint could proceed as planned.

The second Sprint focused on data collection. The majority of data were stored in the department's case management system. In order to make use of them for the project, the data had to undergo a process known as "Extract, Transform, and Load" (ETL). This process allows the end user to construct their own queries on data which has been manipulated into a suitable format (Theodorou et al., 2017).

### 4.5.1 Extract

Initially, a naïve extraction was performed with no filtering. This resulted in every domain in the system being returned. However, not all of the domains were relevant; some had expired, some were unmonitored, and some were test data. Nevertheless, this exercise was useful in order to place an upper-bound on the number of domains. In collaboration with the author's team, the data were filtered down to a more suitable subset.

### 4.5.2 Transform

As demonstrated in Figure 06, DNS is a hierarchical system i.e. in the domain "hmrc.gov.uk", the left-most portion "hmrc" is a sub-domain of "gov.uk". Similarly, "gov" is a subdomain of ".uk".

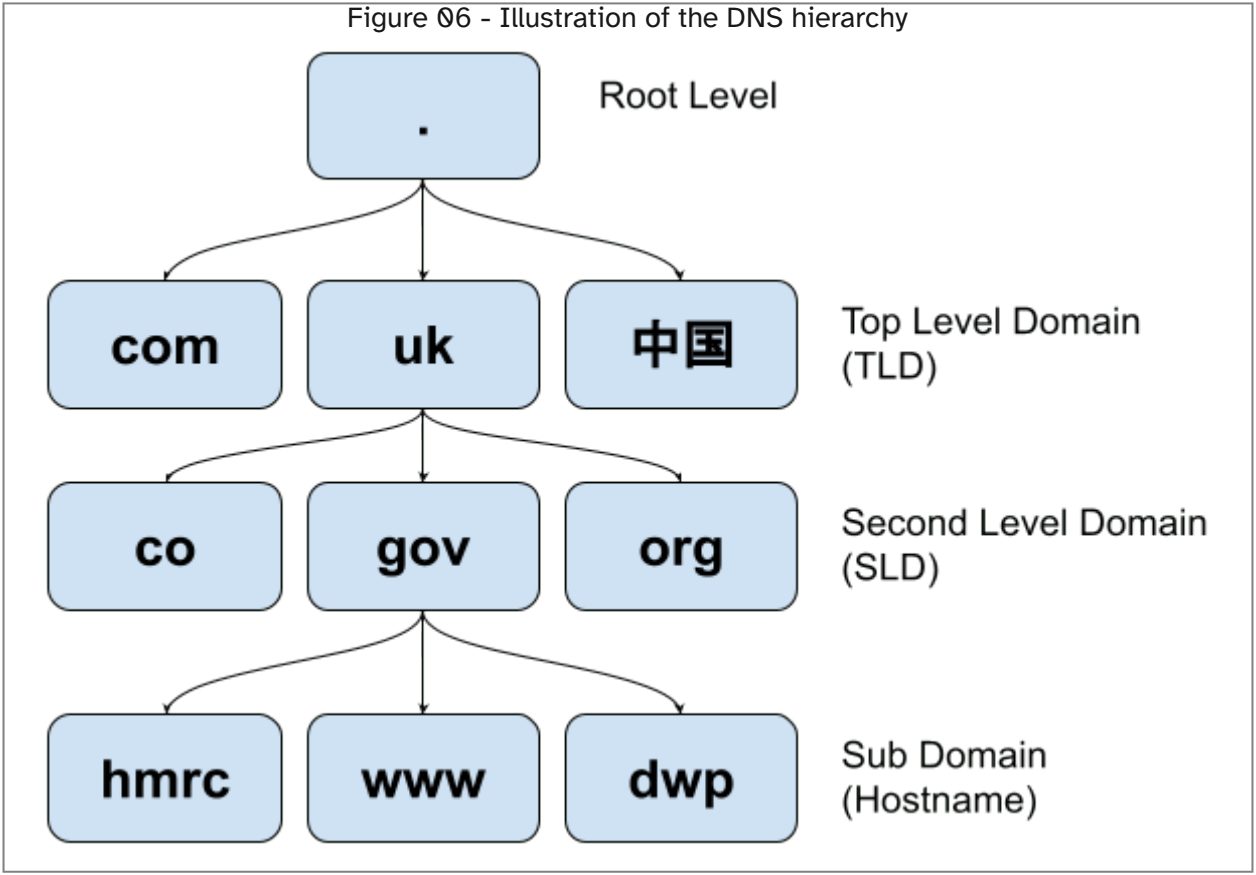

Figure 06 - Illustration of the DNS hierarchy

The data structure for the DAG required that this order be inverted, i.e. formatted as ".uk.gov.hmrc"[1]. A Python program was written to ingest the domain names and reverse their logical order. After testing, it became clear that the DAG required both that there were no duplicates and no logical gaps in the hierarchy. The program was altered to satisfy these requirements. The revised code is in Annex 2 - PoC Source Code. This ability to rapidly react to unexpected issues is a key strength of Agile project management.

4.5.3 Load

A major concern was the performance of the system when loading the data. The unexpurgated dataset consisted of around 300,000 nodes which may have been slow to ingest and render. The system was initially tested with a sample of 4,000 nodes with no noticeable delay. A further test was performed by loading all the data, but truncating the display to the 2nd level. This test also rendered well, as can be seen in Figure 07.

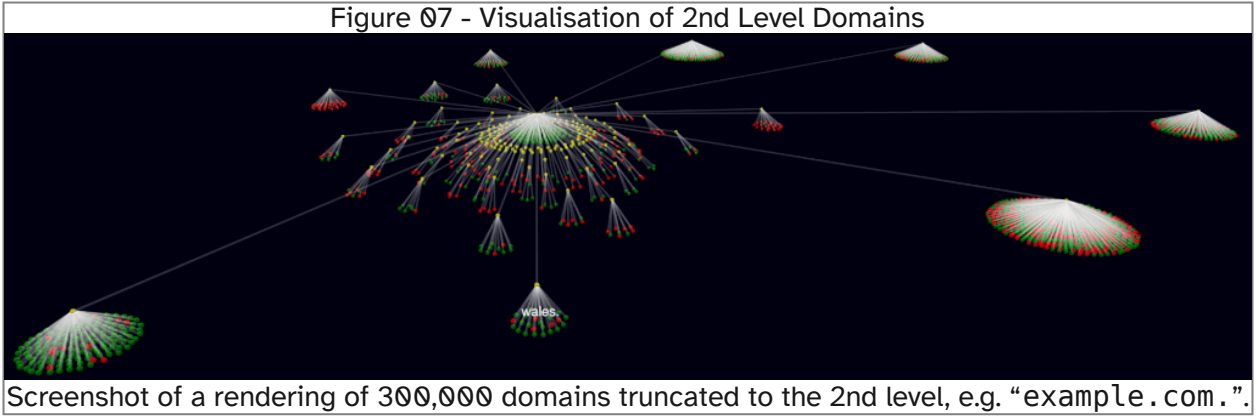

Figure 07 - Visualisation of 2nd Level Domains

Screenshot of a rendering of 300,000 domains truncated to the 2nd level, e.g. "example.com.".

Unfortunately, when displaying the 3rd level, performance decreased dramatically and became unacceptably slow. In order to prevent Cybersickness, it is necessary to maintain a high and consistent frame rate (Weech, Kenny and Barnett-Cowan, 2019). It also became apparent that a large set of domains might not be easily navigable without further work to improve the layout (Figure 08).

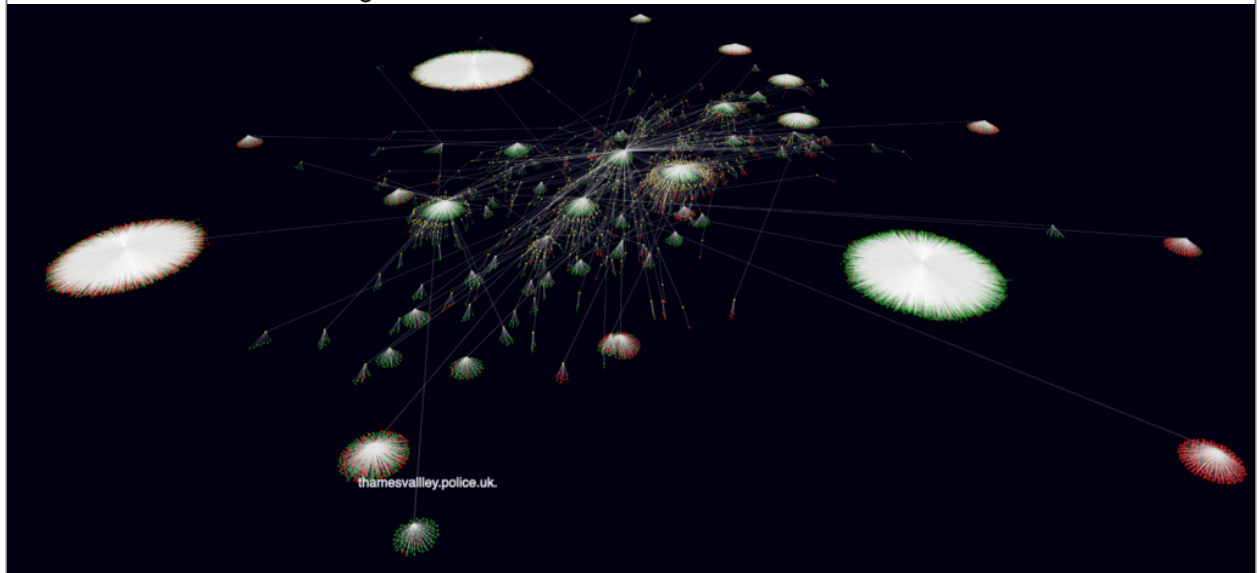

Figure 08 - Visualisation of 3rd Level Domains

A complex rendering of DNS down to the 3rd level, e.g. `thamesvalley.police.uk`

Finally, the PoC showed that it was possible to format and render the data. However, the performance of the platform was insufficient for the volume of data, and the interactive capabilities were non-existent. For the testable prototype, the author moved to a tool which improved performance. The practice of "mind mapping" in VR is actively being explored as a way to help learners organise their thoughts in a hierarchical manner (Sims and Karnik, 2021). The Noda mind-mapping tool (Coding Leap LLC, 2022) can import a formatted CSV and render it as an interactive DAG without performance issues.

An interactive PoC (Figure 09) was constructed using Noda and subsequently became the basis for the tested Metaverse experience.

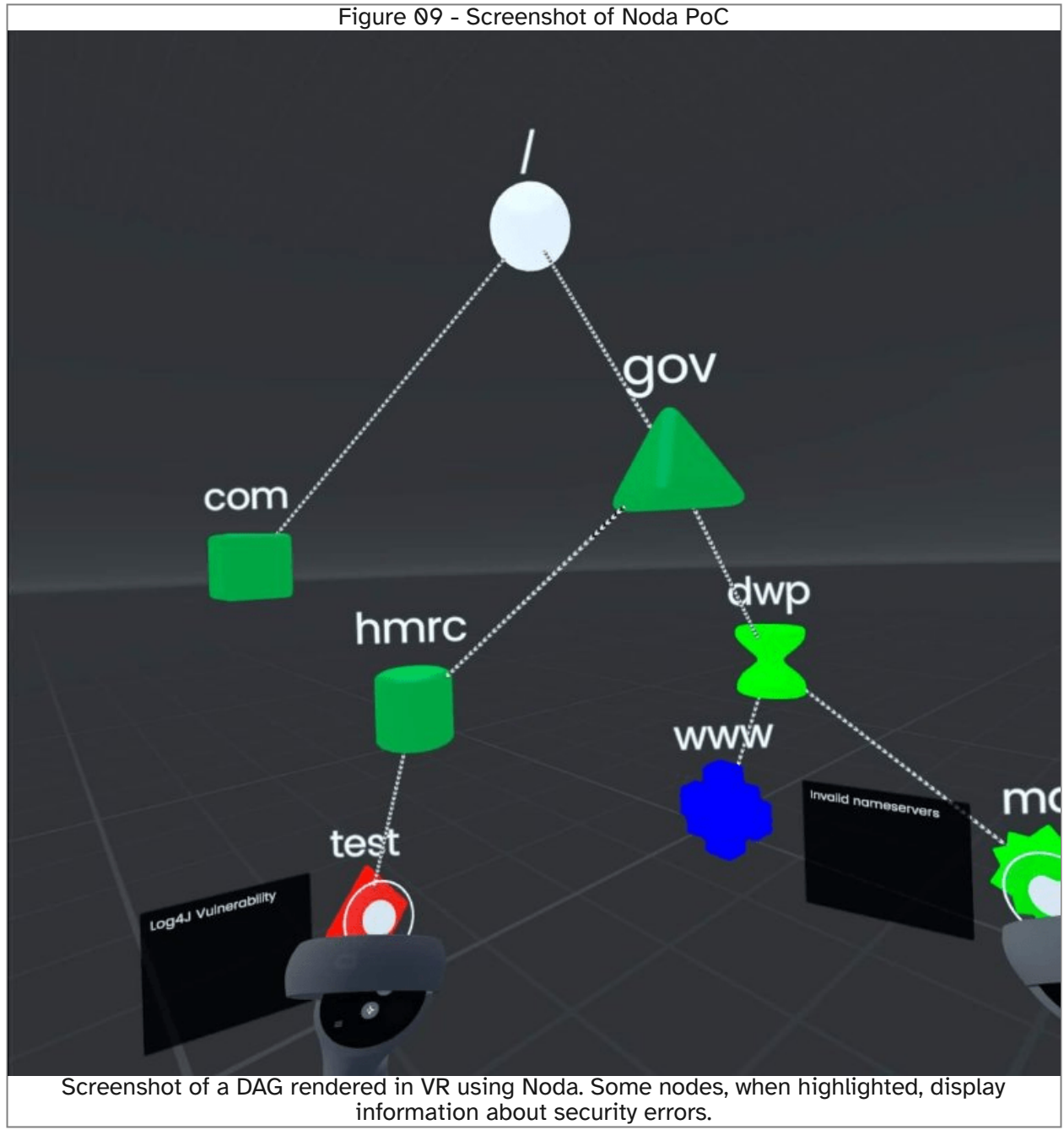

Figure 09 - Screenshot of Noda PoC

Screenshot of a DAG rendered in VR using Noda. Some nodes, when highlighted, display information about security errors.

## 4.6 Construction & Installation

As discussed above, the prototype was constructed in the Noda VR app. Noda ingests a CSV of data and then renders it in the Metaverse. The author constructed an algorithm to transform the extracted data into the format required for Noda. This was then manually altered to add features such as node colouring, shape, and textual description (see Figure 10).

The author considered manually placing the nodes on the 3D canvas. While this is feasible for a small dataset, it is impractical for large datasets. Therefore a second algorithm was used based on the Force-Directed graph layout provided by the popular D3.js library (Poinet, Stefanescu and Papadonikolaki, 2020). This uses the classic Verlet integration to plot the trajectories of multiple nodes, bound by their relationship to descendent and ancestor nodes, but repelled from nodes not within their hierarchy (Verlet, 1967). Both the D3 library and Verlet algorithm are well regarded in the author's industry and are the primary acceptable ways to approach this task (Bostock, Ogievetsky and Heer, 2011).

Once the data were formatted, they were securely transferred to the HMD using USB. The graph was loaded into the environment, manually checked for accuracy, and departmental logos were added. Finally, the environment was saved and backed-up in compliance with the organisation's information retention rules.

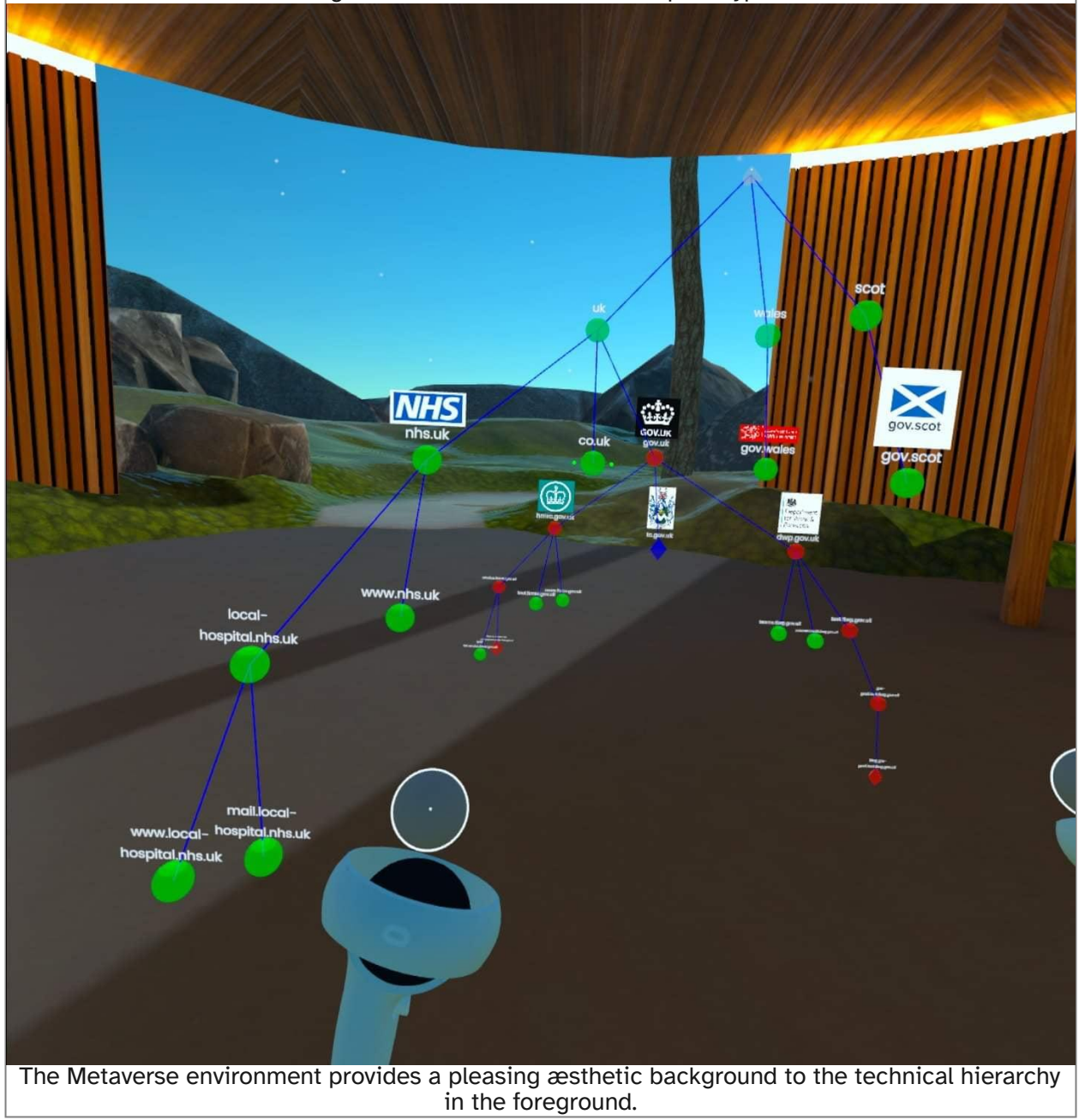

Figure 10 - Screenshot of tested prototype

The Metaverse environment provides a pleasing æsthetic background to the technical hierarchy in the foreground.

## 4.7 Rapport Building

The study consisted of two different groups of participants. The first were members of the author's workplace. They were able to experience the Metaverse by wearing the HMD and interacting with the environment. The second were members of the author's wider industry. They were invited to a video-conference seminar and watched a demonstration video of the Metaverse experience.

A key requirement for user research is for the researcher to build rapport with the subject; this is needed in order to make the subject comfortable, to reassure them that data protection is taken seriously, and to make them feel confident they can discuss with the researcher any issues they encounter. Two different approaches were needed to build rapport with these different groups.

The author examined a number of ways to build rapport, while also delivering mandatory information about consent and Health & Safety. Research has shown that starting an interaction with an impersonal or robotic reading of a prepared statement often makes people uncomfortable (Deakin and Wakefield, 2014). However, the author's industry makes use of templates for building rapport with subjects. These templates have been repeatedly tested and are considered best practice (Calderon, 2022). For the first group, a statement based on these templates was used to build and establish rapport (see Annex 3 - Questionnaire and Results). It is worth noting that because the participants belong to the same team as the author, there was a pre-existing level of rapport.

Although participants in the online study were asked the same questions as the in-person participants, a different approach was needed to build rapport with subjects over video ([Hai-Jew, 2015](#)). To address this, the author delivered a welcoming and informal introduction for their presentation to a cross-Government symposium on data issues. The author encouraged participants to ask questions and provide feedback. Video from this session can be seen in [Annex 4 - Presentation to DataConnect22](#).

## 4.8 Ethical, social and legal considerations

Throughout the project, care was taken to ensure that all modern ethical, social, and legal issues were considered. One aspect which was not fully explored was the negative experience of the Metaverse on users with mobility issues or visual impairments. Research is ongoing to discover how people with visual impairments can fully interact with the Metaverse, but even at this early stage it is apparent that a visual-only Metaverse experience will be discriminatory ([Waskiewicz, 2022](#)). Tackling this issue is beyond the scope of this investigation, but is discussed further in [Chapter 7](#).

## 4.9 Summary

In summary, the Sprints to design the prototype were successful. Data were successfully extracted, transformed, and loaded into a variety of models. The pilot test allowed the author to adjust the parameters of the questionnaire and survey. Finally, user testing and response gathering was undertaken in accordance with the expectations of the author's organisation.

# 5. Results and Analysis

This chapter assesses and analyses the findings from testing the prototype with users. This follows the advice of Bylinskii that "it's not enough to perform a study carefully, it must also be reported thoroughly, so that the reader can understand what was done, what the results show, and how they can be reproduced." ([Bylinskii et al., 2022](#)). The techniques used in this section include product testing with users, statistical surveys, and sentiment analysis.

## 5.1 Questionnaire Construction

A key part of Agile design is testing prototypes with users ([Luojus, Kauppinen and Lahti, 2018](#)). All aspects of the research were designed to be run in an Agile fashion, so both the prototype Metaverse experience and prototype questionnaire were piloted with two participants in order to validate their utility.

The "technology acceptance model" is an established method of assessing users' attitudes towards new technology ([Davis, 1989](#)). The author based their initial questionnaire on a previously designed and tested set of questions from "A Technology Acceptance Model Survey of the Metaverse Prospects" ([Aburbeian, Owda and Owda, 2022](#)). For the questionnaire design, the author considered using an indeterminate Likert Scale for ranking participants' feelings. However, they concluded that a 5-point Likert Scale was more commonly used and would allow participants to quickly complete the questions ([Kandasamy et al., 2020](#)). In order to achieve the best results with a questionnaire, it is important to make sure participants understand the purpose of the study and feel comfortable providing negative feedback. One way to achieve this is to emphasise to the user that it is the service being tested, not the user, and that there are no right or wrong answers ([Bylinskii et al., 2022](#)). This is consistent with the author's employer's practices and was incorporated into the questionnaire instructions given to participants.

As the author has no medical training, they used a pre-existing questionnaire to assess whether participants were suffering from Cybersickness. A variety of questionnaires were considered. Because of its direct applicability to the issue of VR devices, the industry-standard Virtual Reality Sickness Questionnaire (VRSQ) was chosen to assess participants' comfort ([Kim et al., 2018](#)). Kim's version can be seen in [Figure 11](#). It asks participants to self-report any physical issues they experience; including eye-strain, headaches, and vertigo. The author identified a potential weakness of the questionnaire; it fails to address issues with proprioception (also known as kinesthesia) i.e. whether the user feels a disjoint between their perception of their body's movement and reality.

| Figure 11 - Virtual Reality Sickness Questionnaire | | |
|---|---|---|
| VRSQ symptom | Oculomotor | Disorientation |
| 1. General discomfort | O | |
| 2. Fatigue | O | |
| 3. Eyestrain | O | |
| 4. Difficulty focusing | O | |
| 5. Headache | | O |
| 6. Fullness of head | | O |
| 7. Blurred vision | | O |
| 8. Dizzy (eyes closed) | | O |
| 9. Vertigo | | O |
| Total | [1] | [2] |

VRSQ administered to participants after leaving the Metaverse (Kim et al., 2018).

Feedback from the pilot showed that some questions from the VRSQ were ambiguous, for example the term "Fullness of head" was not understood. This is possibly due to translation issues from the original Korean. The questions were adjusted to be clearer for participants (see Annex 3 - Questionnaire). Additionally, feedback from the pilot participants was used to make some æsthetic changes to the design to the prototype and clarify the questionnaire preamble.

## 5.2 Product Testing

Participants who took part in the Metaverse experience were asked to complete a series of tasks. These included navigating to a specific part of the model, reporting on what they saw, and interacting with the data. The full list of tasks can be found in Annex 3 - Questionnaire. This product testing showed that every participant was able to complete each task. All reported that they found the navigation easy to understand and use. Interaction was slightly more difficult and a few participants took some time to familiarise themselves with the controls.

The product testing took place in a private area in the workplace. This was necessary to prevent participants from feeling uncomfortable about performing tasks in public. The nature of the HMD meant that participants could not see their surroundings. Therefore a private space prevented other people from approaching the participants and risking injury to either party.

## 5.3 Results of the Study

As described in Methodology, the study consisted of several steps:

1. Assess participant's familiarity with, and attitudes to, the Metaverse.
2. Help participant put on HMD.
3. Ask participant to complete a series of tasks inside the Metaverse.
4. Administer VRSQ.
5. Ask participant to complete further tasks inside the Metaverse.
6. Remove HMD.
7. Assess participant's attitudes to VR in the workplace.

Participants were divided into two groups. The first group consisted of 8 stakeholder participants from the author's immediate team. They wore the HMD and completed the task in VR, as seen in Figure 12. Testing took place during the working day and in the regular office environment. One potential participant declined to take part in the study, as was their right. The participant explained that they had previously experienced Cybersickness when using a VR HMD and were unwilling to risk their health by taking part. They were happy for their objection to be recorded in the study.

The second group consisted of 45 participants in an online seminar as part of the UK Government's Data Connect Week (GOV.UK, 2022). These participants answered the same questions as the first group, but saw a video presentation of the Metaverse experience, rather than trying it themselves. Because these participants did not enter the Metaverse environment, there was no need to administer the VRSQ, nor ask them about their direct experience of the prototype.

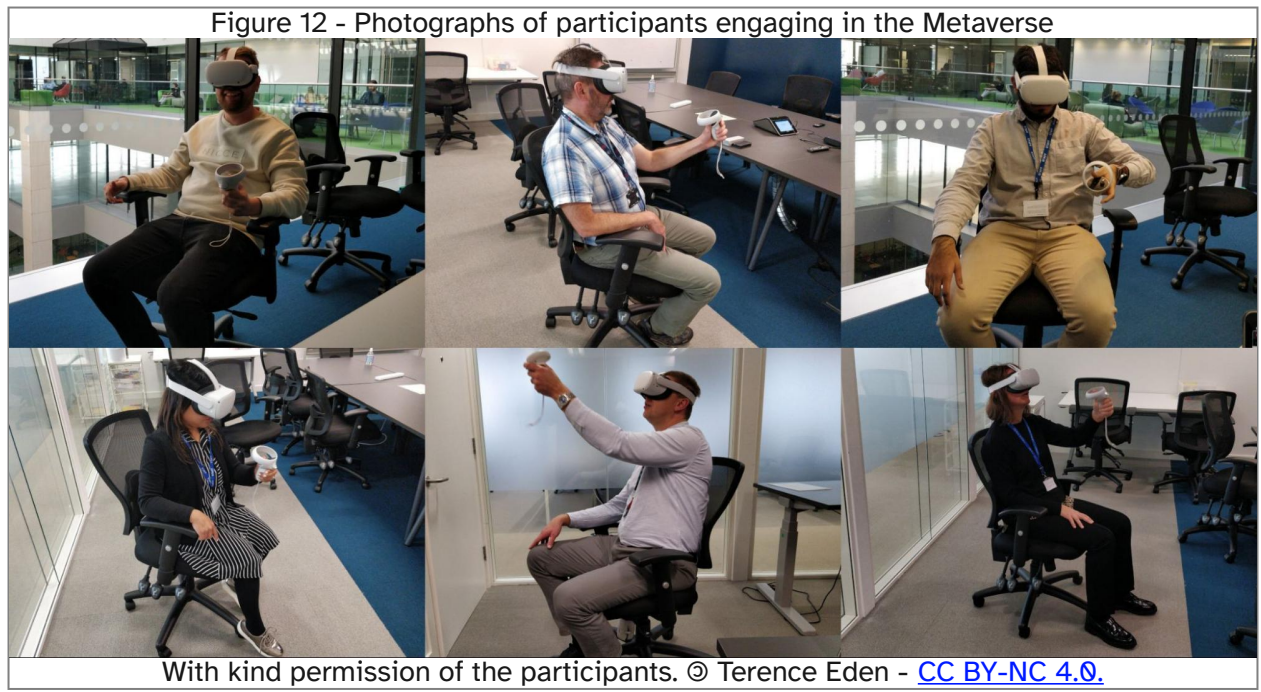

Figure 12 - Photographs of participants engaging in the Metaverse

With kind permission of the participants. © Terence Eden - CC BY-NC 4.0.

Due to the small size of both sample groups, demographic details were not collected as this may have enabled participants to be de-anonymised which would have breached their confidentiality (Narayanan, Huey and Felten, 2016).

## 5.4 Survey Results

Results of the questionnaires were analysed and are presented in summary. Full results can be seen in Annex 3 - Questionnaire and Results.

35% of those attending the online seminar were occasional-to-frequent users of VR, compared to 0% of in-person participants. (Figure 13). This lack of experience may bias the later results as those participants may be more interested in the novelty of the experience rather than the utility.

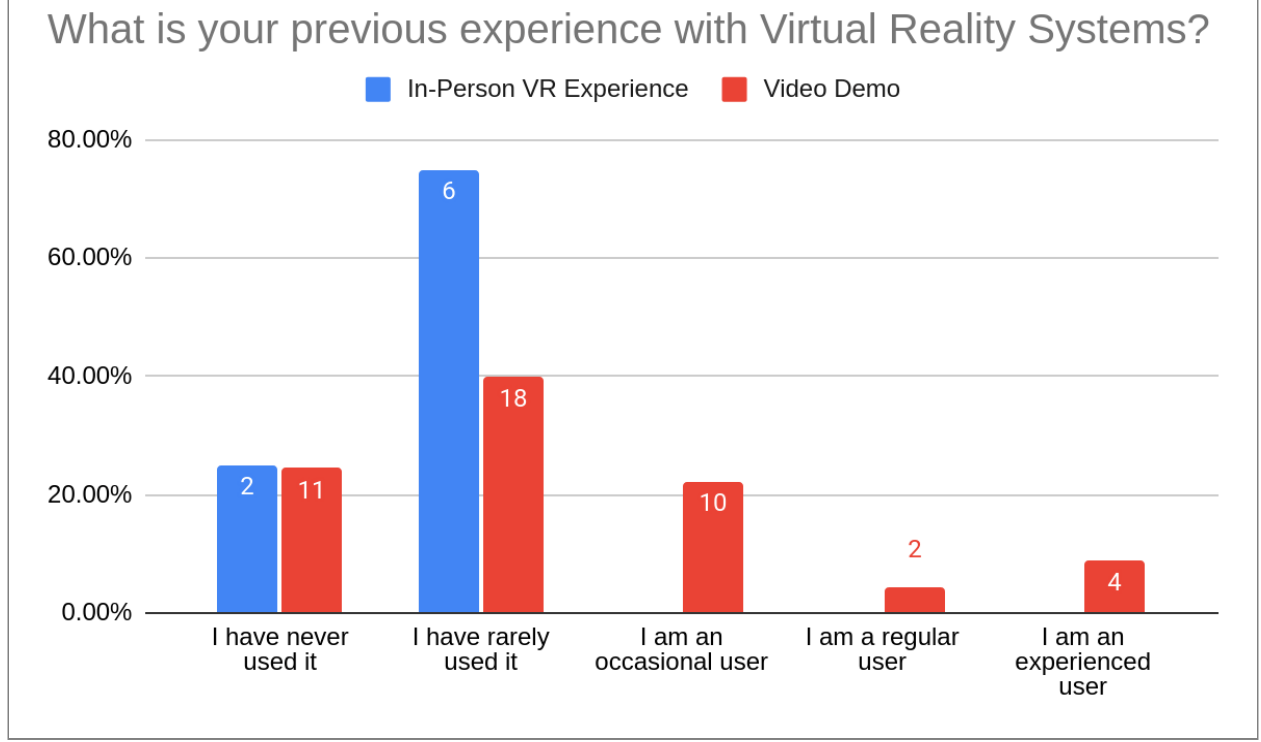

Figure 13 - Questionnaire Results "What is your previous experience with Virtual Reality Systems?"

The participants in the online seminar were self-selecting. This may explain why more of them had experienced the Metaverse compared to the participants in the in-person study who were selected from the author's team.

In contrast, the in-person cohort expressed greater enthusiasm for trying the Metaverse (60%) than those at the seminar (33%).

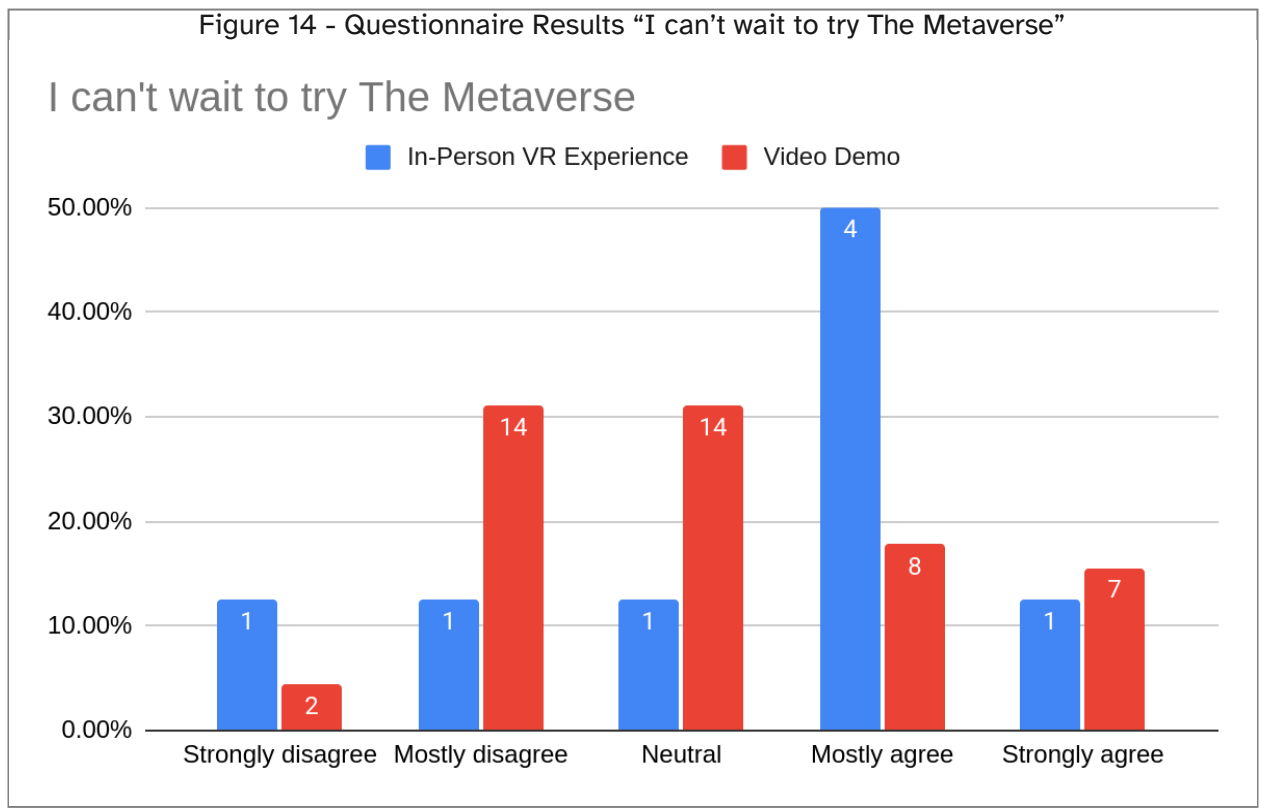

Figure 14 - Questionnaire Results "I can't wait to try The Metaverse"

It is important to note that the in-person participants were all members of the author's team. This personal relationship and lack of anonymity may have caused them to report more favourably so as not to upset or discourage their colleague. This sort of bias is well-known within the user testing community and should be corrected for in a future study (Locascio et al., 2016).

Post-demonstration, the surveys showed there was stronger support for the idea that the Metaverse could be useful to the organisation amongst participants who had directly experienced it (60%) than those who watched the video demo (40%).

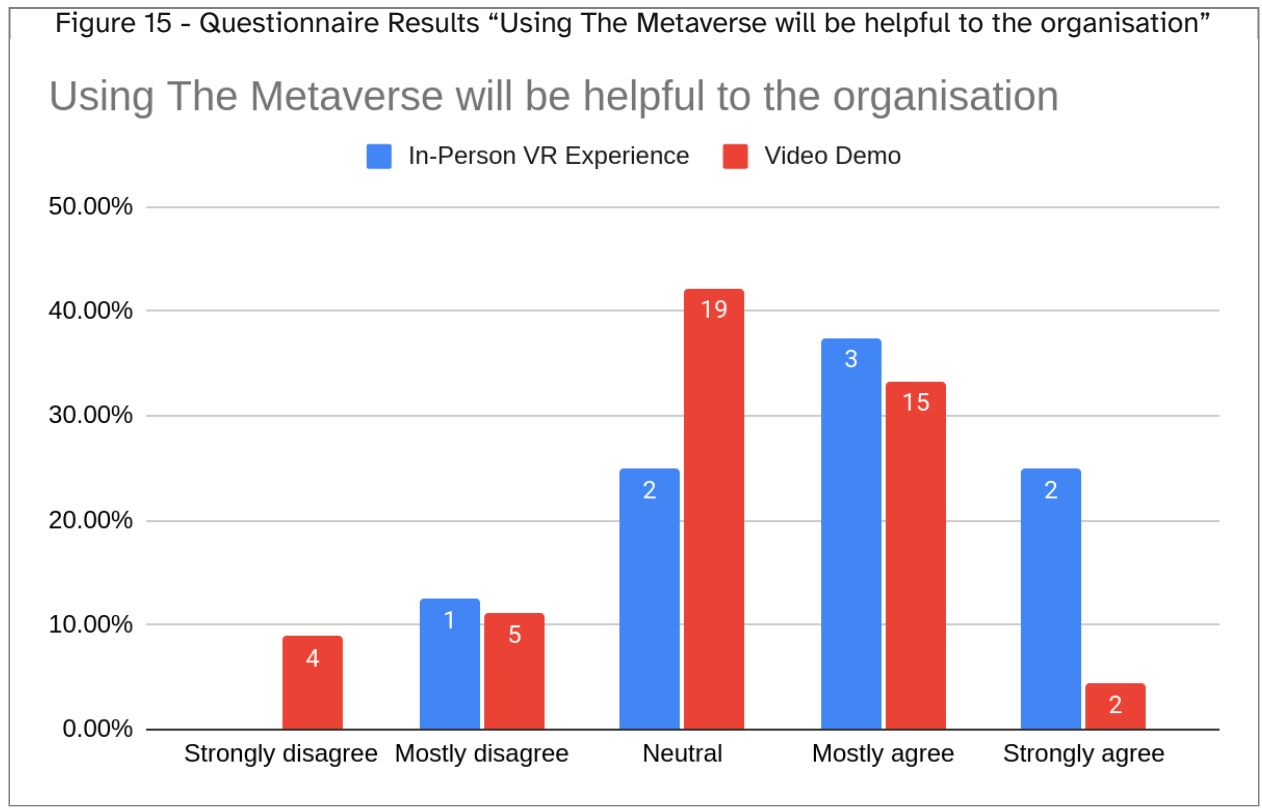

Figure 15 - Questionnaire Results "Using The Metaverse will be helpful to the organisation"

This is perhaps because those who only experienced the video demonstration came from a variety of organisations, not all of which would have data suited to this sort of interactive experience.

Similarly, Figure 16 shows that intention to use the Metaverse was much higher among the in-person cohort (75%) than those who watched the video (40%). Although a minority of both sets of participants expressed reluctance in future use, there was a majority in both sets who supported it.

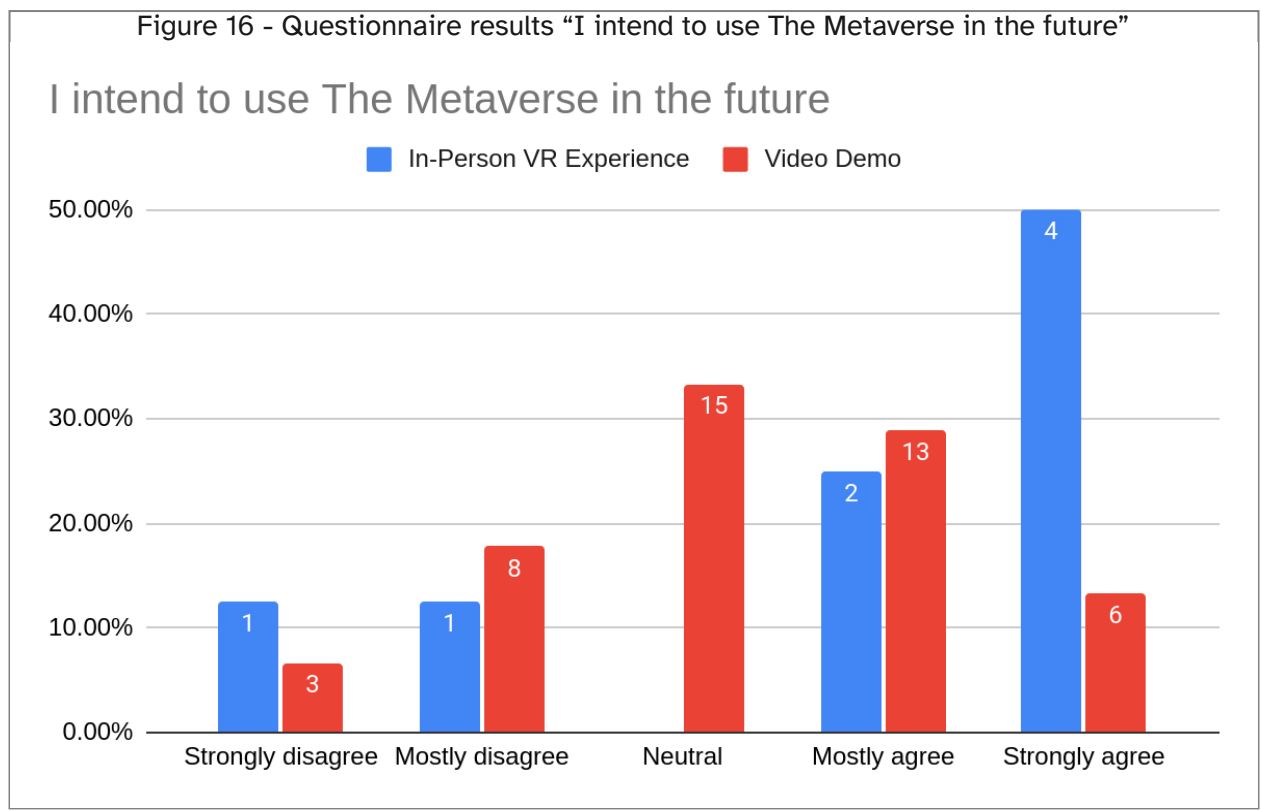

Figure 16 - Questionnaire results "I intend to use The Metaverse in the future"

A short demonstration of a novel experience is not representative of a full working day in the Metaverse. Future studies should use much longer demonstrations to see if that alters users' attitudes towards the experience.

## 5.5 Virtual Reality Sickness Questionnaire

While participants were immersed in the Metaverse, they were asked a series of questions to determine whether they were experiencing Cybersickness. One participant declined to take part citing a previous nausea-inducing experience with VR. Amongst the others, the most common complaint was that the headset was uncomfortable and that the image was blurry. None of the participants experienced nausea, vertigo, eye-strain, or any of the other symptoms listed on the VRSQ. Participants who usually wore eyeglasses were the only ones to mention blurry images. They reported that images were fuzzy, although "No more than I'd expect with my eyesight" and that images were "Fuzzy - like looking at a CRT". This may have been caused by the lack of a "spacer" to adjust the focal length of the lenses ([Botha, de Wet and Botma, 2021](#)).

The only other symptoms mentioned were due to the HMD. One participant reported "throbbing in the back of the head. Because it is so tight" and three specifically mentioned the heaviness of the headset. One participant mentioned that they had a brief dizzy spell after they had removed the HMD. These reactions are consistent with the effects both predicted and observed in the literature (Hawkinson, 2022). The discomfort was spontaneously mentioned by several participants before the VRSQ was administered, meaning it was readily apparent to users without prompting.

This lack of Cybersickness could be attributable to the relatively short amount of time participants spent in VR. It may also be because the prototype was optimised to ensure a smooth frame-rate. It is believed that the more realistic the VR environment is, the worse the symptoms of Cybersickness are ([Tiiro, 2018](#)). The Noda environment, while sufficiently detailed to give the illusion of complexity, is not a photorealistic simulation (see [Figure 10](#) and [Annex 4](#)). This may have reduced the risk of Cybersickness. However, there is a great deal of variation between individuals' susceptibility to Cybersickness (Rebenitsch and Owen, 2014). It may be the case that the cohort had a low predisposition to a negative reaction. It may also be the case that participants were unwilling to reveal medical information or admit their distress.

## 5.6 Selected comments

Both sets of participants were able to give free-text commentary to each question. This allowed them to be more expressive about their feelings than is provided for by the restrictive Likert scale. While it is possible to perform automated sentiment analysis and opinion mining using neural networks and Bayesian classifiers ([Li et al., 2019](#)) there was only a small amount of data, so these responses were manually sorted by sentiment. This generated an emotional map of participants' feelings towards the Metaverse. The volume of data gathered from the written responses roughly corresponds to those of the Likert scale responses, but they provide much more nuance, detail, and insight into the strength of those feelings.

As discussed in the [Literature Review](#), the term "Metaverse" has been used in this paper as a generic term for VR worlds. However, the comments from participants show that they inexorably link the term with the Meta Corporation (still referred to by its old name, "Facebook", by all respondents).

|  | Selected participant comments from questions asked before the demonstration | | |
| --- | --- | --- | --- |
| Questions | Negative | Positive | Author's Analysis |
| What is your opinion on The Metaverse? | "I guess it's because it's becoming synonymous with Facebook, and… well… Facebook."<br><br>"Haven't tried it but everything I've seen about it has been negative."<br><br>"trying to monopolise on the VR industry - it's not good." | "Positive if there is an agreement on what it actually is. Seems to be being thrown around as the latest buzzword right now"<br><br>"Has potential, still not found a killer app yet though" | The negative sentiment is mostly focussed on Facebook and participants' dislike and distrust of the company.<br><br>Any positive sentiment is tempered by the fact that it is seen as having high potential rather than actual usefulness. |
| I can't wait to try The Metaverse | "Suspicious of data gathering activities (eg Meta)"<br><br>"Whatever the metaverse can do a normal computer application could do. It sounds boring. It's a gimmick"<br><br>"There are no good headsets of a doable value." | "I'm a doer not a reader - being able to see it with my own eyes will be amazing. I'm just really excited."<br><br>"Sounds interesting and a new way of seeing and understanding data" | Participants were eager to try the Metaverse overall, but doubts remain especially over price of headsets and the potential for corporate abuse. |

|  | Selected participant comments from questions asked after the demonstration | | |
| --- | --- | --- | --- |
| Questions | Negative | Positive | Author's Analysis |
| The Metaverse experience is exciting. | "Devalues what it means to be human, too many risks as humans are inherently evil"<br><br>"it doesn't feel much different to previous attempts at 3d visualisations" | "Level of interaction you can't get anywhere else."<br><br>"A different way of seeing things. Instead of seeing things in a text format - I could see visually and go into depth with the things that you wanted."<br><br>"I can see it being a good tool to visualise complex issues" | There is ample positive sentiment toward the Metaverse, but undercut with an acknowledgement that 3D experiences have been tried before without success. |
| Using The Metaverse will be helpful to the organisation. | "large cash outlay"<br><br>"anything interactive I've seen (eg Zuckerberg demos) is far short of anything useful"<br><br>"This could be done on a flat screen."<br><br>"my concern is that this could turn into a case of Doing Metaverse Because Everyone Is Doing Metaverse, and ultimately trying to shoehorn it into applications where it isn't the most appropriate tool" | "I'm a visual learner. Being able to jump into a diagram made it easier to understand and more interesting."<br><br>"Our dataset is complex. Analysts have trouble navigating it. Using the metaverse would help them"<br><br>"Provides new ways to illustrate structures/problems that the organisation is facing."<br><br>"turning highly complex data into something a human can interact with is amazing" | The major concerns are around the cost of hardware, and that the hype around the Metaverse will lead to unsustainable products.<br><br>Again, participants see high value in being able to interact with complex data sets and demonstrate their complexity to others. |
| I intend to use The Metaverse in the future | "I doubt it would be practical or possible to implement within the civil service"<br><br>"All I think about is Facebook. When I hear metaverse all I hear is an extension of facebook. And I have no interest in that."<br><br>"They are looking to get people addicted and monetised. It's not coming from a place where they want to help humanity - it is just business expansion. Especially when this could be quite addictive." | "Lots of potential for collaborative work, spatial placing etc."<br><br>"It would be beneficial for a range of people to understand the scale and complexity of the data that my organisation works with"<br><br>"Metaverse is huge and has many different uses and scope. For me and my job I can see the benefit"<br><br>"It's just natural. It's part of evolution to act in virtual spaces." | Once users had either seen a demonstration, or experienced the Metaverse, their sentiment became much more positive.<br><br>This was tempered by the realisation that the cost was likely to be prohibitive.<br><br>There is also a prevailing attitude that corporate IT departments in the public sector are a blocker to innovation (Smith, 2013). |
| Any further thoughts on what you've seen? | "Does Facebook get my data?"<br><br>"I'm reluctant because of what I've heard. Environments you go in and | "really brings the idea to life in a way that a static slide could never do."<br><br>"The more we can empower people to | The Metaverse experience is seen as empowering and interesting. But, once again, Facebook's reputation, and the negative press |
| | there are no rules and people come and attack you."<br><br>"Having removed the headset - I felt it was a bit heavy." | create their own visualisations, the more powerful this tool will be"<br><br>"Definitely something I want to try again. An awesome piece of tech."<br><br>"I didn't feel silly using it." | associated with it, are barriers to adoption. |

Some users are obviously concerned about the direction in which this technology is going, which may lead to them rejecting the journey into the Metaverse. Research has shown that some people believe all humans are "biologically and psychologically maladapted to life in a technological society", leading to those people feeling alienated from modern life ([Fleming, 2022](#)).

The analysis of the survey shows that the testing of the prototype was successful. It also quantifies the appetite for use of the Metaverse in the workplace. The results of the VRSQ did not raise significant concerns about the health effects of VR in short sessions. These results need to be caveated with the recognition that this was a small study with a specific user group. This critical evaluation is covered in the next section.

# 6. Critical Evaluation of Results

The aim of this research was to build a prototype to see if users could interact with complex hierarchical Cybersecurity data which had been rendered in the Metaverse, and to assess users' perceptions of both the experience and the Metaverse in a business context. It also sought to explore reported health issues with wearing VR headsets.

These aims were successfully achieved. Custom code was created to extract, transform, and load the data from existing systems, through industry-standard algorithms, and into a new environment. Participants shared a range of opinions about using VR in the organisation. While there was some scepticism, users were mostly supportive of adopting the Metaverse for interactive data. However the phrase "Metaverse" was repeatedly met with hostility from participants, which stemmed from the association with Facebook. The health impacts of using HMDs were explored and some limitations were found.

## 6.1 Health and Wellbeing Issues

Previous research has shown that one of the limitations of using an HMD is that users find them uncomfortable ([Mehrfard et al., 2019](#)) as demonstrated in [Figure 17](#). While work is ongoing to make headsets smaller, lighter, and more comfortable, the current state-of-the-art is relatively primitive. This constrains the acceptance of HMD hardware among potential users.

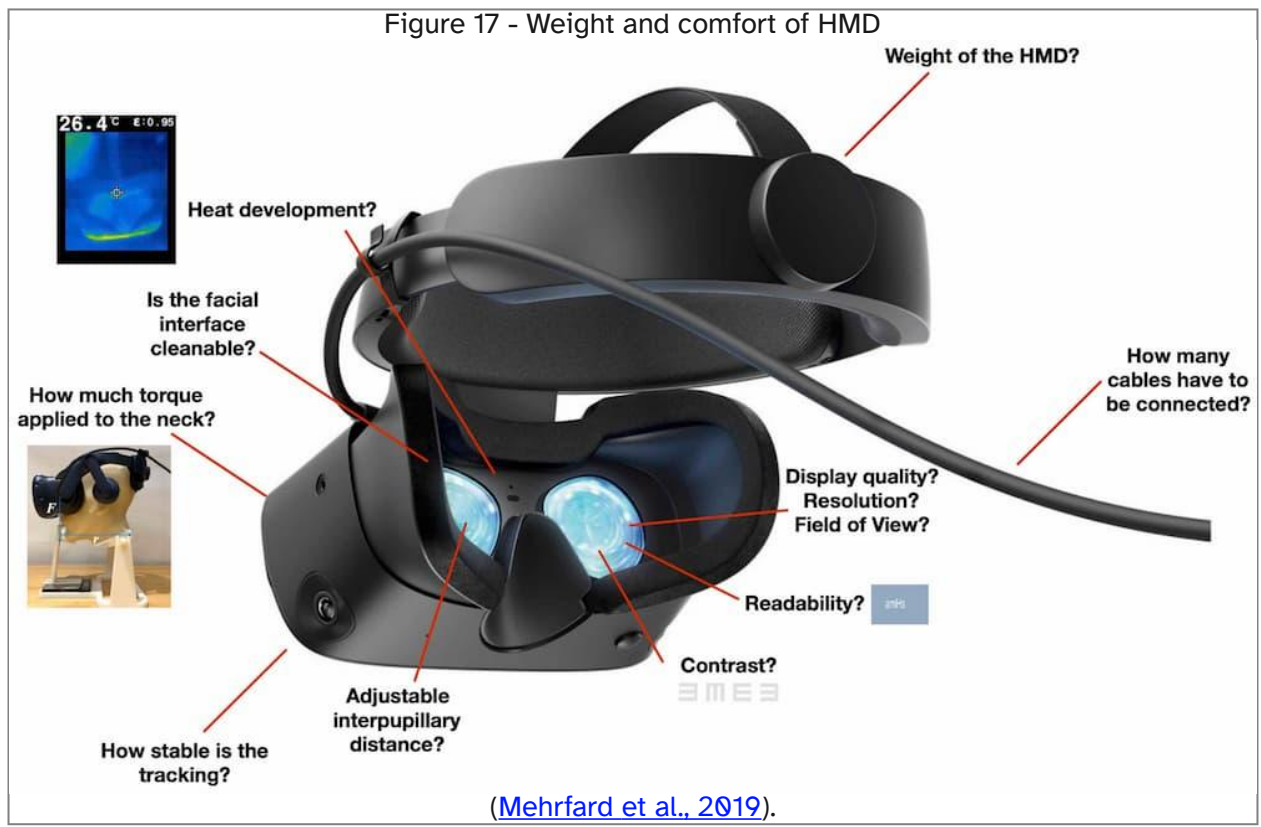

Figure 17 - Weight and comfort of HMD
([Mehrfard et al., 2019](#)).

All of the issues raised in Mehrfard et al. (2019) were replicated in this study - along with a few additional concerns. For example, the foam padding of the HMD is known to cause allergic reactions in some people which caused Meta to include a silicone cover with later revisions of their hardware (['Oculus recalls Quest 2 headset pads after skin rashes', 2021](#)). During this study, the silicone cover was prone to slipping off the HMD and blocking its lenses. Some participants found wearing the device to be claustrophobic and were worried about not seeing their surroundings or people approaching them. Concern was also expressed at the high cost of the HMD and its apparent fragility; one participant was worried about damaging the expensive equipment. These are psychological concerns as opposed to physiological, but indicate possible causes of resistance to VR adoption in the workplace.

An unexpected finding was that setting up the HMD for each participant required a lengthy calibration process. Multiple factors can have an effect on the comfort of the headset and the clarity of image. For example, InterPupillary Distance (IPD) varies considerably between humans and a mismatch between the user's IPD and the IPD setting on the HMD can cause discomfort ([Kim et al., 2021](#)). Additionally, differing cranium sizes result in a need for different tightness settings on the HMD head strap. Finally, the position and comfort of the head strap was impacted by participants' various hairstyles.

All these factors may have contributed to the negative sentiment expressed by some participants. Their wellbeing was negatively impacted by their frustration at the poor clarity of the images, the discomfort of wearing the HMD, and the lengthy calibration routine.

## 6.2 Data Presentation Issues

When evaluating the success of a new method of presenting data, a popular metric comes from Edward Tufte who asked the chart designer to consider the "data-ink ratio" ([Tufte, 2001](#)). This is often referred to by the pithy epithet "show the data". Tufte encourages creators of visualisations to eschew superfluous decoration and to concentrate on the presentation of data. While Tufte's work has undeniably been influential, his theories are all based on the qualitative feedback of one subject - himself. A later quantitative study showed that, in contrast to Tufte's preferences, some people prefer graphs which displayed a low data-ink ratio ([Inbar, Tractinsky and Meyer, 2007](#)).

In order to assess the prototype, a Tuftean analysis of the VR environment was conducted by the author and is reported in [Table 03](#). It shows that the use of VR to display complex information meets many, but not all, of Tufte's recommendations.

| Table 03 - How did the prototype compare to "Tufte's instructions to practitioners" ([Tufte, 2001](#))? | |
| --- | --- |
| Show the data | All the data are shown. |
| Induce the viewer to think about the substance rather than about methodology, graphic design, the technology of graphic production, or something else | Participants spent a lot of time "enjoying the view" and marvelling at the environment. Once the shock of the new wore off, they then engaged with the substance. |
| Avoid distorting what the data have to say | The data were plainly arranged. There was no distortion or manipulation. |
| Present many numbers in a small space | Space is less of a concern in VR as it has an endless topography. Nevertheless, a large amount of data were presented in a relatively compact form. |
| Make large data sets coherent | The ability to collapse large data sets into smaller, more manageable chunks makes this more coherent. |
| Encourage the eye to compare different pieces of data | The use of colour and shape allowed participants to easily compare data. |
| Reveal the data at several levels of detail, from a broad overview to the fine structure | The interactive nature of the structure gives both an overview and fine-grained detail. |
| Serve a reasonably clear purpose: description, exploration, tabulation, or decoration | The purpose of the visualisation was exploration. However, some elements (such as website logos) were purely decorative. |
| Be closely integrated with the statistical and verbal descriptions of a data set | The prototype is separate from any statistical or verbal descriptions. |

## 6.3 Industry Applicability

The applicability of the study to the wider industry may be limited. The prototype was only tested on a small cohort who were already familiar with the data. They were all members of the author's team, which may have affected their objectivity. Although demographic data was not recorded on a participant level, the author's team does not have sufficient gender diversity to be able to draw robust conclusions about whether there is a notable difference in acceptance between genders. Similarly, employees at the author's workplace do not represent a wide range of ages. Therefore it isn't possible to assess whether the perception of the Metaverse differs by age.

The prototype only tested highly-structured hierarchical data. Participants were able to navigate and interact with data in this new paradigm, but not all organisations have data which is curated and structured. This type of navigation may be restricted to datasets which are both massive and well-defined.

Cost is a significant barrier to adoption. The UK Government wishes to "maximise efficiency within budgets" in order to save money ([Sunak, 2022](#)). Despite the ability of technology to improve the efficiency and impact of the Civil Service, there is a reluctance among senior politicians to fully embrace new ways of working ([Smith, 2021](#)). Against this backdrop, it is uncertain whether the organisation will be prepared to invest in the technology necessary for widespread Metaverse deployment.

## 6.4 Current Academic Research

This research builds on recent work which demonstrates the need for effective visualisations of Cybersecurity data ([Kullman and Engel, 2022](#)). While 3D visualisations of complex data are not new ([Berkel and Bos, 1999](#)), the price of hardware has fallen and the speed of software has improved to such a degree that this research shows that it is finally feasible to use VR in the workplace to explore complex data sets.

Contemporary research shows the promotional hype around the Metaverse doesn't necessarily match the enthusiasm of potential users ([Dwivedi et al., 2022](#)). These findings are replicated in this study. Participants mostly enjoyed the immersive VR experience, but their comments highlighted several drawbacks to the technology which undermine its supposed utility.

## 6.5 Reflection on research

The author reflects on the research using the DIEP reflective model ([Rogers, 2001](#)).

- Describe
    - A prototype was constructed and tested. Results from participants were gathered and evaluated.
- Interpret
    - A technology literate workforce may be receptive to engaging with VR for certain tasks. However, the author's workplace may not be typical of all organisations.
- Evaluate effectiveness
    - The project was a success. Participants mostly reacted positively to it.
- Plan for the future
    - Engage with the organisation to refine and develop the prototype. Perform further research to establish the utility and accessibility of the Metaverse.

## 6.6 Ethical, legal, and social issues

Participants' data were processed according to GDPR. The Official Secrets Act was observed in relation to the secret data being used. The study did not provide a way for people with visual impairments to interact with the prototype. This is a recognised limitation with the current state of VR ([Seigneur and Choukou, 2022](#)). While users who experience Cybersickness could be accommodated through the use of 2D video demonstrations, there was no effective method of including those for whom VR is inaccessible. Deployment of this technology, without reasonable adjustments, would likely be a breach of the Equality Act.

## 6.7 Conclusion

Testing the prototype showed that it is possible for users to interact with familiar hierarchical data in the Metaverse. Participants mostly found the experience both useful and compelling, but it should be acknowledged that the novelty of using VR in the workplace may have had an impact on their reactions. The name "Metaverse", and the reputation of Facebook, are barriers to adoption. Finally, although the majority of participants didn't suffer any of the effects of Cybersickness, the HMDs were uncomfortable to wear even for a short period.

There is further work to be done to make the Metaverse experience more useful for organisations, and this will be discussed in the next section.

# 7. Further Work, Conclusions, and Recommendations

## 7.1 Strength and presentation of findings

This was a small study, performed on people familiar with the existing data. Nevertheless, the study shows that participants exhibited interest and excitement about bringing the Metaverse experience into the workplace. This is tempered by concern about the company which controls it.

This paper started by defining [three main aims](#); build a prototype to visualise complex information in the Metaverse, assess professionals' attitudes to using VR in the workplace, and investigate the health constraints of using VR equipment.

The prototype worked well from a technology perspective and, judging from the comments of participants, was a useful way to visualise the data and show the scale of issues faced. The findings indicate that the processing power of modern HMDs is insufficient for models with extremely large volumes of data (see [Load](#)). However, when using smaller models, the fidelity and fluidity of the Metaverse environment was sufficient to create an experience which was both useful and compelling.

The survey showed participants had both an acceptance of VR in the workplace and a willingness to find ways to exploit this technology. While some doubts remain over the suitability of Meta to be an effective steward of the fledgling Metaverse, sentiment towards the technology's use in the workplace was broadly positive.

Participants expressed some health concerns with the HMDs, particularly around the weight and comfort of the device. While no users complained of eye-strain, some reported that the images were blurry. Widespread Cybersickness did not occur in this study, although one potential participant refused to take part having previously experienced VR-induced nausea.

## 7.2 Recommendations

The author recommends that the term "Metaverse" should not be used in the generic sense due to its inescapable link with Facebook. Instead, a more neutral term should be adopted, for example "Virtual Reality".

Technology companies frequently misappropriate terms from science fiction in order to make their products sound innovative and exciting ([Newitz, 2021](#)). This trope is illustrated in [Figure 18](#). The original Metaverse envisioned by Stephenson was a dystopian nightmare. It served as an allegory for how the worst excesses of humanity can corrupt an environment. Given this, the author finds it puzzling that Facebook chose to rebrand their company around the notion of the Metaverse. Perhaps this behaviour should not be surprising as several major technology projects have already been named after dystopian fictional products such as Skynet, Palantir, and Panopticon ([Levendowski, 2022](#)).

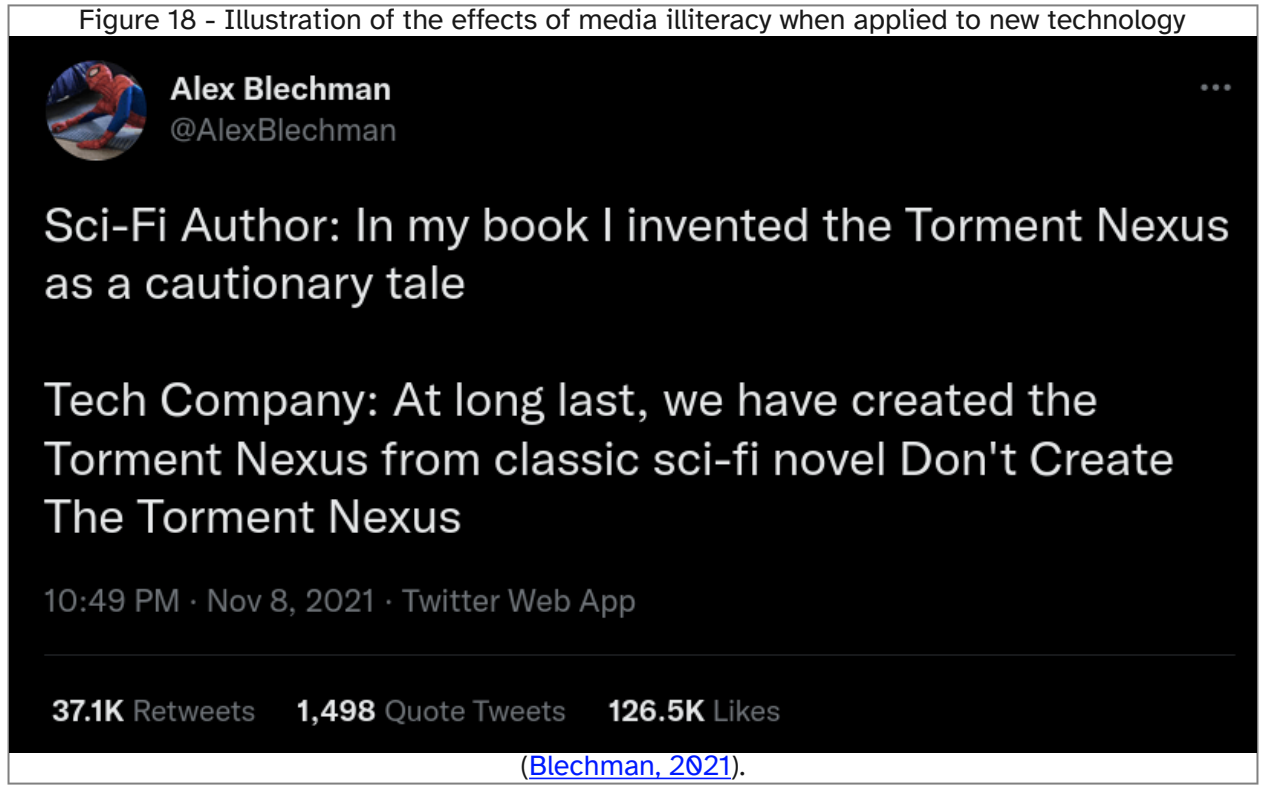

Figure 18 - Illustration of the effects of media illiteracy when applied to new technology ([Blechman, 2021](#)).

The author contends that companies cannot escape the reputational damage which accompanies these choices. Several participants brought up the science-fiction roots of the name "Metaverse" and the poor reputation of the "Meta" company. This should be taken as a warning signal that there remains a strong emotional barrier to adopting technologies promoted by platforms with questionable workplace practices ([Chitkara, 2020](#)).

## 7.3 Further research

Employers may be unwilling to adopt technology which opens them up to claims of workplace discrimination. Currently, people with visual impairments are excluded from participating in Virtual Reality, therefore accessibility research needs to be prioritised. Similarly, ongoing work into the cause of Cybersickness and its prevention will be necessary to ensure these products are usable by all employees.

## 7.4 Suggestions for improved practice

Users face significant privacy risks when interacting with VR ([Nair, Garrido and Song, 2022](#)) - a concern noted by some participants. It is imperative that future solutions are designed in such a way as to preserve users' privacy and dignity. With an increase in workplace monitoring ([TUC, 2020](#)) it seems inevitable that some employers will abuse the Metaverse to intrude on their employees' personal data. Best practice guidelines should be developed to ensure that business tools in the Metaverse prevent this sort of intrusion. Further investigation is needed to determine how best to warn users of the risks they face from the Metaverse.

In addition, employers need to have a better understanding of Cybersickness ([Garcia-Agundez et al., 2019](#)). Intense use of HMDs may not be suitable for a significant number of employees. Given that up to 90% of users already experience eye-strain from the use of digital devices ([Coles-brennan, Sulley and Young, 2019](#)), it is likely that users of HMDs will also experience these issues.

In the author's opinion, HMDs cannot become a shared resource. Today, employees waste considerable time configuring shared computer equipment to meet their personal needs ([Smith Brain Trust, 2019](#)). The author previously noted the long and convoluted set-up time required to adjust an HMD to fit each participant. Making employees reconfigure a shared HMD for each use would add an unacceptable time burden. With shared HMDs being a potent vector for antibiotic-resistant bacteria and other pathogens, the risk of disease transfer is also high ([Creel et al., 2020](#)). Therefore, any organisation intending to use the Metaverse should ensure that it has sufficient budget to provide every user with their own HMD and also supply suitable cleaning products.

## 7.5 Further Exploitation Within The Organisation

Participants undoubtedly found the prospect of using the Metaverse exciting, and some saw how it could be useful in the future. But this is tempered with four main drawbacks:

1. The cost of Metaverse devices is high and may be more than the organisation is willing to spend on experimental technology. It is not feasible to share equipment due to health & safety concerns and because each user requires bespoke adjustments to the HMD.
2. Lack of accessibility. Employers will need to provide reasonable adjustments for visually impaired users and for those suffering from Cybersickness.
3. Physiological factors. The current generation of HMDs are bulky, heavy, hot, and difficult to adjust. Adoption may have to wait until a majority of these concerns have been addressed.
4. Political concerns. It may be unpalatable to expend resources on something which might be seen as frivolous. This is a pernicious problem in Government innovation spending - commonly referred to by the pejorative term "Daily Mail Test" ([Gearson et al., 2020](#)).

## 7.6 Conclusion

This small study shows the viability of using the Metaverse in the workplace to examine complex hierarchical data. While concerns remain about the health impacts of wearing Virtual Reality equipment, users seem excited to explore the potential of the Metaverse. It remains to be seen whether the negative reputation of the Metaverse will be a significant barrier to wider adoption.

There is a long history of researchers creating new and baroque methods to visualise hierarchical information. For example the complexity of the early World Wide Web led to the development of Hyperbolic Trees as a navigation tool ([Lamping and Rao, 1996](#)). However, despite the technical proficiency and innovative approach of Hyperbolic navigation, it did not become mainstream. The first modern VR display was created in the middle of the last century ([Bradley, 1965](#)) but did not become a commercial success.

The use of 3D environments helps solve some problems associated with investigating large amounts of data but may not be easily integrated into existing workflows. Although VR is currently enjoying a resurgence of interest via the Metaverse, it is uncertain whether it will become widely adopted.

# References

Aburbeian, A.M., Owda, A.Y. and Owda, M. (2022) 'A Technology Acceptance Model Survey of the Metaverse Prospects', AI, 3(2), pp. 285–302. Available at: https://doi.org/10.3390/ai3020018.

Anderson, R.J. (2011) 'Florence Nightingale: The Biostatistician', Molecular Interventions, 11(2), pp. 63–71. Available at: https://doi.org/10.1124/mi.11.2.1.

Ball, M. (2022) The metaverse: And how it will revolutionize everything. First edition. New York, NY: Liveright Publishing Corporation, a division of W.W. Norton & Company.

Berdanier, C.G.P. and Lenart, J.B. (2021) So, you have to write a literature review a guided workbook for engineers. Available at: https://ieeexplore.ieee.org/servlet/opac?bknumber=9214308 (Accessed: 25 June 2022).

Berkel, B. van and Bos, C. (1999) Move. Amsterdam: UN Studio & Goose Press.

Bernstein, S.L., Weiss, J. and Curry, L. (2020) 'Visualizing implementation: Contextual and organizational support mapping of stakeholders (COSMOS)', Implementation Science Communications, 1(1), p. 48. Available at: https://doi.org/10.1186/s43058-020-00030-8.

Blechman, A. (2021) 'Sci-Fi Author: In my book I invented the Torment Nexus as a cautionary tale Tech Company: At long last, we have created the Torment Nexus from classic sci-fi novel Don't Create The Torment Nexus', Twitter. Available at: https://twitter.com/AlexBlechman/status/1457842724128833538 (Accessed: 29 August 2022).

Bostock, M., Ogievetsky, V. and Heer, J. (2011) 'D³ Data-Driven Documents', IEEE Transactions on Visualization and Computer Graphics, 17(12), pp. 2301–2309. Available at: https://doi.org/10.1109/TVCG.2011.185.

Botha, B.S., de Wet, L. and Botma, Y. (2021) 'Experts' review of a virtual environment for virtual clinical simulation in South Africa', Computer Animation and Virtual Worlds, 32(2), p. e1983. Available at: https://doi.org/10.1002/cav.1983.

Bradley, W.E. (1965) 'Remotely controlled remote viewing system'. Available at: https://patents.google.com/patent/US3205303A/en (Accessed: 23 July 2022).

British Computer Society (2021) 'BCS Code of Conduct | BCS'. Available at: https://www.bcs.org/membership/become-a-member/bcs-code-of-conduct/ (Accessed: 14 November 2021).

Bylinskii, Z. et al. (2022) 'Towards Better User Studies in Computer Graphics and Vision'. Available at: https://doi.org/10.48550/ARXIV.2206.11461.

Calderon, H. (2022) 'Sharing our new user research templates and guides - Service Transformation', ESSEX.GOV.UK. Available at: https://servicetransformation.blog.essex.gov.uk/2022/03/17/sharing-our-new-user-research-templates-and-guides/ (Accessed: 21 September 2022).

Castleberry, A. and Nolen, A. (2018) 'Thematic analysis of qualitative research data: Is it as easy as it sounds?', Currents in Pharmacy Teaching and Learning, 10(6), pp. 807–815. Available at: https://doi.org/10.1016/j.cptl.2018.03.019.

Chitkara, H. (2020) 'Facebook Workplace's abandoned 'content control' tools highlight the limitations of remote work', Business Insider. Available at: https://www.businessinsider.com/facebook-drops-censorship-tools-for-workplace-2020-6 (Accessed: 14 October 2022).

Coding Leap LLC (2022) 'Noda - mind mapping in virtual reality, solo or group'. Available at: https://noda.io/ (Accessed: 11 September 2022).

Coles-brennan, C., Sulley, A. and Young, G. (2019) 'Management of digital eye strain', Clinical and Experimental Optometry, 102(1), pp. 18–29. Available at: https://doi.org/10.1111/cxo.12798.

Creel, B. et al. (2020) 'Bacterial Load of Virtual Reality Headsets', in 26th ACM Symposium on Virtual Reality Software and Technology. New York, NY, USA: Association for Computing Machinery (VRST '20), pp. 1–8. Available at: https://doi.org/10.1145/3385956.3418958.

Davis, F.D. (1989) 'Perceived Usefulness, Perceived Ease of Use, and User Acceptance of Information Technology', MIS Quarterly, 13(3), p. 319. Available at: https://doi.org/10.2307/249008.

Deakin, H. and Wakefield, K. (2014) 'Skype interviewing: Reflections of two PhD researchers', Qualitative Research, 14(5), pp. 603–616. Available at: https://doi.org/10.1177/1468794113488126.

Dionisio, J.D.N., III, W.G.B. and Gilbert, R. (2013) '3D Virtual worlds and the metaverse: Current status and future possibilities', ACM Computing Surveys, 45(3), pp. 34:1–34:38. Available at: https://doi.org/10.1145/2480741.2480751.

Dwivedi, Y.K. et al. (2022) 'Metaverse beyond the hype: Multidisciplinary perspectives on emerging challenges, opportunities, and agenda for research, practice and policy', International Journal of Information Management, 66, p. 102542. Available at: https://doi.org/10.1016/j.ijinfomgt.2022.102542.

Egliston, B. and Carter, M. (2021) 'Critical questions for Facebook's virtual reality: Data, power and the metaverse', Internet Policy Review, 10(4). Available at: https://doi.org/10.14763/2021.4.1610.

Fleming, S. (2022) 'The Unabomber and the origins of anti-tech radicalism', Journal of Political Ideologies, 27(2), pp. 207–225. Available at: https://doi.org/10.1080/13569317.2021.1921940.

Fuertes, V. (2021) 'The rationale for embedding ethics and public value in public administration programmes', Teaching Public Administration, 39(3), pp. 252–269. Available at: https://doi.org/10.1177/01447394211028275.

Garcia-Agundez, A. et al. (2019) 'Development of a Classifier to Determine Factors Causing Cybersickness in Virtual Reality Environments', Games for Health Journal, 8(6), pp. 439–444. Available at: https://doi.org/10.1089/g4h.2019.0045.

Gearson, J. et al. (2020) Whole Force By Design: Optimising Defence to Meet Future Challenges. Available at: https://www.sercoinstitute.com/research/2020/whole-force-by-design-optimising-defence-to-meet-future-challenges (Accessed: 21 September 2022).

Gleeson, P. (2017) Working with coders: A guide to software development for the perplexed non-techie. New York, NY: Apress.

GOV.UK (2021) 'National AI Strategy', GOV.UK. Available at: https://www.gov.uk/government/publications/national-ai-strategy/national-ai-strategy-html-version (Accessed: 15 October 2022).

GOV.UK (2022) 'DataConnect22 is coming!', GOV.UK. Available at: https://www.gov.uk/government/news/dataconnect22-is-coming (Accessed: 12 October 2022).

Grazier, K.R. and Cass, S. (2017) 'Let's Get Digital: Computers in Cinema', in K.R. Grazier and S. Cass (eds) Hollyweird Science: The Next Generation: From Spaceships to Microchips. Cham: Springer International Publishing (Science and Fiction), pp. 131–183. Available at: https://doi.org/10.1007/978-3-319-54215-7_5.

Hai-Jew, S. (ed.) (2015) Enhancing qualitative and mixed methods research with technology. Hershey, PA: Business Science Reference, an imprint of IGI Global (Advances in knowledge acquisition, transfer, and management (AKATM) book series).

Hamshere, J.D. and Blakemore, M.J. (1976) 'Computerizing Domesday Book', Area, 8(4), pp. 289–294. Available at: https://www.jstor.org/stable/20001145 (Accessed: 20 July 2022).

Harrop, W. and Armitage, G. (2006) 'Real-time collaborative network monitoring and control using 3D game engines for representation and interaction', in Proceedings of the 3rd international workshop on Visualization for computer security. New York, NY, USA: Association for Computing Machinery (VizSEC '06), pp. 31–40. Available at: https://doi.org/10.1145/1179576.1179583.

Heide, M. (2022) 'From Threat to Risk: Changing Rationales and Practices of Secrecy', Public Integrity, 24(3), pp. 254–266. Available at: https://doi.org/10.1080/10999922.2021.1932144.

Hoffman, P.E., Sullivan, A. and Fujiwara, K. (2019) DNS Terminology. Request for Comments RFC 8499. Internet Engineering Task Force. Available at: https://doi.org/10.17487/RFC8499.

Inbar, O., Tractinsky, N. and Meyer, J. (2007) 'Minimalism in information visualization: Attitudes towards maximizing the data-ink ratio', in Proceedings of the 14th European conference on Cognitive ergonomics: Invent! explore! New York, NY, USA: Association for Computing Machinery (ECCE '07), pp. 185–188. Available at: https://doi.org/10.1145/1362550.1362587.

Kandasamy, I. et al. (2020) 'Indeterminate Likert scale: Feedback based on neutrosophy, its distance measures and clustering algorithm', Soft Computing, 24(10), pp. 7459–7468. Available at: https://doi.org/10.1007/s00500-019-04372-x.

Kim, H.K. et al. (2018) 'Virtual reality sickness questionnaire (VRSQ): Motion sickness measurement index in a virtual reality environment', Applied Ergonomics, 69, pp. 66–73. Available at: https://doi.org/10.1016/j.apergo.2017.12.016.

Kim, J.-S. et al. (2021) 'Estimation of Interpupillary Distance Based on Eye Movements in Virtual Reality Devices', IEEE Access, 9, pp. 155576–155583. Available at: https://doi.org/10.1109/ACCESS.2021.3128991.

Kohnke, A. (2020) 'The Risk and Rewards of Enterprise Use of Augmented Reality and Virtual Reality', ISACA [Preprint]. Available at: https://www.isaca.org/resources/isaca-journal/issues/2020/volume-1/the-risk-and-rewards-of-enterprise-use-of-augmented-reality-and-virtual-reality (Accessed: 28 August 2022).

Kullman, K. and Engel, D. (2022) 'Interactive Stereoscopically Perceivable Multidimensional Data Visualizations for Cybersecurity', Journal of Defence & Security Technologies, 4(1), pp. 37–52. Available at: https://doi.org/10.46713/jdst.004.03.

Lamping, J. and Rao, R. (1996) 'Visualizing large trees using the hyperbolic browser', in Conference Companion on Human Factors in Computing Systems. New York, NY, USA: Association for Computing Machinery (CHI '96), pp. 388–389. Available at: https://doi.org/10.1145/257089.257389.

Lee, H., Woo, D. and Yu, S. (2022) 'Virtual Reality Metaverse System Supplementing Remote Education Methods: Based on Aircraft Maintenance Simulation', Applied Sciences, 12(5), p. 2667. Available at: https://doi.org/10.3390/app12052667.

Levendowski, A. (2022) 'Dystopian Trademark Revelations'. Rochester, NY. Available at: https://papers.ssrn.com/abstract=4231409 (Accessed: 28 September 2022).

Li, Z. et al. (2019) 'A survey on sentiment analysis and opinion mining for social multimedia', Multimedia Tools and Applications, 78(6), pp. 6939–6967. Available at: https://doi.org/10.1007/s11042-018-6445-z.

Locascio, J. et al. (2016) 'Utilizing Employees as Usability Participants: Exploring When and When Not to Leverage Your Coworkers', in Proceedings of the 2016 CHI Conference on Human Factors in Computing Systems. New York, NY, USA: Association for Computing Machinery (CHI '16), pp. 4533–4537. Available at: https://doi.org/10.1145/2858036.2858047.

Lopez, J. and Chisholm, A. (2021) 'The Civil Service Apprenticeship Strategy 2021-2022', GOV.UK. Available at: https://www.gov.uk/government/publications/the-civil-service-apprenticeship-strategy-2021-2022 (Accessed: 23 October 2022).

Luojus, S., Kauppinen, S. and Lahti, J. (2018) 'Developing Higher Education: Agile Methods in Service Design', in 11th annual International Conference of Education, Research and Innovation. Seville, Spain, pp. 4540–4548. Available at: https://doi.org/10.21125/iceri.2018.2022.

Mehrfard, A. et al. (2019) 'A Comparative Analysis of Virtual Reality Head-Mounted Display Systems'. arXiv. Available at: https://doi.org/10.48550/arXiv.1912.02913.

Moore, N. et al. (2021) 'Innovation During a Pandemic: Developing a Guideline for Infection Prevention and Control to Support Education Through Virtual Reality', Frontiers in Digital Health, 3. Available at: https://www.frontiersin.org/articles/10.3389/fdgth.2021.628452 (Accessed: 12 September 2022).

Morse, J.M. (2002) 'A Comment on Comments', Qualitative Health Research, 12(1), pp. 3–4. Available at: https://doi.org/10.1177/1049732302012001001.

Nair, V., Garrido, G.M. and Song, D. (2022) 'Exploring the Unprecedented Privacy Risks of the Metaverse'. Available at: https://doi.org/10.48550/ARXIV.2207.13176.

Narayanan, A., Huey, J. and Felten, E.W. (2016) 'A Precautionary Approach to Big Data Privacy', in S. Gutwirth, R. Leenes, and P. De Hert (eds) Data Protection on the Move. Dordrecht: Springer Netherlands, pp. 357–385. Available at: https://doi.org/10.1007/978-94-017-7376-8_13.

Newitz, A. (2021) 'Why tech companies don't get science fiction', New Scientist, 252(3362), p. 26. Available at: https://doi.org/10.1016/S0262-4079(21)02120-5.

Northumbria University (2022) 'Ethics & Integrity'. Available at: https://www.northumbria.ac.uk/research/ethics-and-integrity/ (Accessed: 28 August 2022).

'Oculus recalls Quest 2 headset pads after skin rashes' (2021) BBC News [Preprint]. Available at: https://www.bbc.com/news/technology-57997112 (Accessed: 18 September 2022).

'Official Secrets Act' (1989). Statute Law Database. Available at: https://www.legislation.gov.uk/ukpga/1989/6/contents (Accessed: 29 August 2021).

Oliver, A. et al. (2019) 'VR Macintosh Museum: Case Study of a WebVR Application', in Á. Rocha et al. (eds) New Knowledge in Information Systems and Technologies. Cham: Springer International Publishing (Advances in Intelligent Systems and Computing), pp. 275–284. Available at: https://doi.org/10.1007/978-3-030-16184-2_27.

Ozkan, N. (2019) 'Imperfections Underlying the Manifesto for Agile Software Development', in 2019 1st International Informatics and Software Engineering Conference (UBMYK), pp. 1–6. Available at: https://doi.org/10.1109/UBMYK48245.2019.8965504.

Parr, T.J. and Rohaly, T.F. (1995) 'A language for creating and manipulating VRML', in Proceedings of the first symposium on Virtual reality modeling language - VRML '95. San Diego, California, United States: ACM Press, pp. 123–131. Available at: https://doi.org/10.1145/217306.217323.

Pearlman, D.M. and Gates, N.A. (2010) 'Hosting Business Meetings and Special Events in Virtual Worlds: A Fad or the Future?', Journal of Convention & Event Tourism, 11(4), pp. 247–265. Available at: https://doi.org/10.1080/15470148.2010.530535.

Petri, K. et al. (2020) 'Effects of Age, Gender, Familiarity with the Content, and Exposure Time on Cybersickness in Immersive Head-mounted Display Based Virtual Reality', American Journal of Biomedical Sciences, pp. 107–121. Available at: https://doi.org/10.5099/aj200200107.

Petricioli, L. and Fertalj, K. (2022) 'Agile Software Development Methods and Hybridization Possibilities Beyond Scrumban', in 2022 45th Jubilee International Convention on Information, Communication and Electronic Technology (MIPRO), pp. 1093–1098. Available at: https://doi.org/10.23919/MIPRO55190.2022.9803402.

Poinet, P., Stefanescu, D. and Papadonikolaki, E. (2020) 'SpeckleViz: A Web-based Interactive Activity Network Diagram for AEC', SimAUD 2020: Proceedings of the 11th annual Symposium on Simulation for Architecture and Urban Design (SimAUD). San Diego, CA, USA: The Society for Modeling and Simulation International (SCS). Available at: http://simaud.org/2020/proceedings/42.pdf (Accessed: 21 September 2022).

Rogers, R.R. (2001) 'Reflection in Higher Education: A Concept Analysis', Innovative Higher Education, 26(1), pp. 37–57. Available at: https://doi.org/10.1023/A:1010986404527.

Runciman, B. (2011) Leaders in computing changing the digital world. Swindon, U.K.: British Informatics Society Ltd. Available at: http://site.ebrary.com/id/10582847 (Accessed: 13 August 2022).

Scalia, P. et al. (2022) 'Eliciting patients' healthcare goals and concerns: Do questions influence responses?', Chronic Illness, 18(3), pp. 708–716. Available at: https://doi.org/10.1177/17423953211067417.

Schwind, V. et al. (2019) 'Using Presence Questionnaires in Virtual Reality', in Proceedings of the 2019 CHI Conference on Human Factors in Computing Systems. New York, NY, USA: Association for Computing Machinery (CHI '19), pp. 1–12. Available at: https://doi.org/10.1145/3290605.3300590.

Seigneur, J.-M. and Choukou, M.-A. (2022) 'How should metaverse augment humans with disabilities?', in 13th Augmented Human International Conference. New York, NY, USA: Association for Computing Machinery (AH2022), pp. 1–6. Available at: https://doi.org/10.1145/3532525.3532534.

Shalf, J. (2020) 'The future of computing beyond Moore's Law', Philosophical Transactions of the Royal Society A: Mathematical, Physical and Engineering Sciences, 378(2166), p. 20190061. Available at: https://doi.org/10.1098/rsta.2019.0061.

Shen, W.-C.M., Lee, C.-C. and Wang, T.-W. (2020) 'Potential bias in creative chart design: A review of nontraditional financial graphs in corporate annual reports', Interactions, 28(1), pp. 58–65. Available at: https://doi.org/10.1145/3434569.

Sims, R. and Karnik, A. (2021) 'VERITAS: Mind-Mapping in Virtual Reality', in 2021 7th International Conference of the Immersive Learning Research Network (iLRN). Eureka, CA, USA: IEEE, pp. 1–8. Available at: https://doi.org/10.23919/iLRN52045.2021.9459348.

Smith Brain Trust (2019) 'Does It Seem Like Everyone Hates Hot-Desking? (Yes) | Maryland Smith', University of Maryland. Available at: https://www.rhsmith.umd.edu/research/does-it-seem-everyone-hates-hot-desking-yes (Accessed: 2 October 2022).

Smith, J. (2021) 'COVID-19, Brexit and the United Kingdom – a year of uncertainty', The Round Table, 110(1), pp. 62–75. Available at: https://doi.org/10.1080/00358533.2021.1875686.

South, L. et al. (2022) 'Effective Use of Likert Scales in Visualization Evaluations: A Systematic Review', Computer Graphics Forum, 41(3), pp. 43–55. Available at: https://doi.org/10.1111/cgf.14521.

Spielberg, S. et al. (1993) 'Jurassic Park'. US. Available at: https://ui.eidr.org/view/content?id=10.5240/1534-FF7C-1702-2A95-6D72-W.

Stirling, M. (2018) 'Perceptions of Hothouse Earth: Science As Advertorial'. Rochester, NY. Available at: https://doi.org/10.2139/ssrn.3243151.

Sunak, R. (2022) 'A message from Prime Minister Rishi Sunak'. Available at: https://us7.campaign-archive.com/?u=69b9426e2863a15f9cfa0a407\&id=1e8add48a3 (Accessed: 9 November 2022).

Taherdoost, H. (2019) 'What Is the Best Response Scale for Survey and Questionnaire Design; Review of Different Lengths of Rating Scale / Attitude Scale / Likert Scale'. Rochester, NY. Available at: https://papers.ssrn.com/abstract=3588604 (Accessed: 4 September 2022).

Tanlamai, U., Savetpanuvong, P. and Kunarittipol, W. (2011) 'Mixed Reality Visualization of Financial Accounting Data', Journal of Information Technology Applications and Management, 18(1), pp. 1–14. Available at: https://doi.org/10.21219/jitam.2011.18.1.001.

Taylor, S. and Soneji, S. (2022) 'Bioinformatics and the Metaverse: Are We Ready?', Frontiers in Bioinformatics, 2. Available at: https://www.frontiersin.org/articles/10.3389/fbinf.2022.863676 (Accessed: 23 July 2022).

Tenny, S. et al. (2022) 'Qualitative Study', in StatPearls. Treasure Island (FL): StatPearls Publishing. Available at: http://www.ncbi.nlm.nih.gov/books/NBK470395/ (Accessed: 24 July 2022).

Theodorou, V. et al. (2017) 'Frequent patterns in ETL workflows: An empirical approach', Data & Knowledge Engineering, 112, pp. 1–16. Available at: https://doi.org/10.1016/j.datak.2017.08.004.

Tiiro, A. (2018) 'Effect of visual realism on cybersickness in virtual reality'. Available at: https://www.academia.edu/en/65563780/Effect_of_visual_realism_on_cybersickness_in_virtual_reality (Accessed: 2 October 2022).

Timonen, T. et al. (2021) 'Virtual reality improves the accuracy of simulated preoperative planning in temporal bones: A feasibility and validation study', European Archives of Oto-Rhino-Laryngology, 278(8), pp. 2795–2806. Available at: https://doi.org/10.1007/s00405-020-06360-6.

TUC (2020) Technology managing people - The worker experience. Available at: https://www.tuc.org.uk/research-analysis/reports/technology-managing-people-worker-experience (Accessed: 28 August 2022).

Tufte, E.R. (2001) The visual display of quantitative information. 2nd ed. Cheshire, Conn: Graphics Press.

Verlet, L. (1967) 'Computer "Experiments" on Classical Fluids. I. Thermodynamical Properties of Lennard-Jones Molecules', Physical Review, 159(1), pp. 98–103. Available at: https://doi.org/10.1103/PhysRev.159.98.

Viire, E. (1997) 'Health and safety issues for VR', Communications of the ACM, 40(8), pp. 40–41. Available at: https://doi.org/10.1145/257874.257882.

Vitillo, A. (2021) 'This is not a tweet, it is a #metaverse message… #VirtualReality #VR #AR https://t.co/f9aDJ2mbc3', Twitter. Available at: https://twitter.com/SkarredGhost/status/1419399248433623040 (Accessed: 20 August 2022).

Walliman, N. (2010) Research Methods: The Basics. Zeroth. Routledge. Available at: https://doi.org/10.4324/9780203836071.

Warner, N. and Teo, J.T. (2021) 'Neurological injury from virtual reality mishap', BMJ Case Reports CP, 14(10), p. e243424. Available at: https://doi.org/10.1136/bcr-2021-243424.

Waskiewicz, M.Z. (2022) 'Virtual Reality for the Visually Impaired'. Available at: https://bora.uib.no/bora-xmlui/handle/11250/3002298 (Accessed: 21 September 2022).

Weech, S., Kenny, S. and Barnett-Cowan, M. (2019) 'Presence and Cybersickness in Virtual Reality Are Negatively Related: A Review', Frontiers in Psychology, 10, p. 158. Available at: https://doi.org/10.3389/fpsyg.2019.00158.

Woodin, G., Winter, B. and Padilla, L. (2022) 'Conceptual Metaphor and Graphical Convention Influence the Interpretation of Line Graphs', IEEE Transactions on Visualization and Computer Graphics, 28(2), pp. 1209–1221. Available at: https://doi.org/10.1109/TVCG.2021.3088343.

Yang, S.H. and Nam, C. (2018) 'What do Consumers Prefer for the Attributes of Virtual Reality Head-mount Displays', in. Calgary: International Telecommunications Society (ITS). Available at: https://www.econstor.eu/handle/10419/184971 (Accessed: 23 November 2022).

Zuill, W. (2012) 'The 8 Agile Maxims of Woody Zuill', Life, Liberty, and the Pursuit of Agility. Available at: http://zuill.us/WoodyZuill/2012/09/16/the-8-agile-maxims-of-woody-zuill/ (Accessed: 20 August 2022).

Zyda, M. (2022) 'Let's Rename Everything "the Metaverse!"', Computer, 55(3), pp. 124–129. Available at: https://doi.org/10.1109/MC.2021.3130480.

# Annex

This annex consists of four pieces of evidence relating to the project.

## Annex 1 - Risk Assessment

Based on the [Health and Safety Executive's template](#).

| What are the hazards? | Who might be harmed and how? | What are you already doing to control the risks? | What further action do you need to take to control the risks? | Who needs to carry out the action? |
|---|---|---|---|---|
| Cybersickness | Participants may experience nausea or other symptoms. | Administering a VRSQ. Informing the participant they can take a break from the experiment at any time. | Ensure cleaning supplies are available during the experiment. | Investigator |
| Display screen equipment | Participants risk eye-strain, posture problems, repetitive strain injuries from overuse or improper use of the HMD and controllers. | Ensure sessions are short. Discuss risks with participants. | Tell participants to report any injury or illness. | Investigator |
| Collision and impact | Due to the HMD obstructing the participant's view, they may walk into walls or impact their extremities against obstacles. | Remove obstacles from the room. Remind participants to stay seated. | Monitor participants to make sure they do not injure themselves. | Investigator |
| Electrical | Both participant and investigator could suffer electric shocks from using faulty HMD equipment. | HMD to be regularly inspected to ensure it is not damaged. Faulty equipment to be replaced. | PAT test electrical equipment connected to the main electrical supply. | Investigator |
| COVID & other biohazards | Participants may leak biological fluids onto devices and spread disease | HMD to use silicone cover on porous areas. | Clean HMD with biocide wipes after every use. | Investigator |

More information on managing risk: www.hse.gov.uk/simple-health-safety/risk

# Annex 2 - PoC Source Code

HTML & JavaScript source code to the PoC used to assess the feasibility of the project.

```html
<!DOCTYPE html>
<head>
        <title>VR DAG from CSV</title>

        <script src="js/d3-dsv.js"></script>
        <script src="js/3d-force-graph-vr.js"></script>

        <style> body { margin: 0; } </style>
        <style>
                .clickable { cursor: unset !important }
        </style>
</head>

<body>
        <div id="3d-graph"></div>

        <script>
                //      Root ID - may change. Possibly load dynamically?
                const rootId = "/";

                //      Load the CSV
                fetch('dns.csv')
                        .then(r => r.text())
                        .then(d3.csvParse)
                        .then(gData =>{
                                //      Set up the arrays
                                const nodes = [], links = [], childLinks = [];

                                //      Loop through each line of the CSV
                                gData.forEach( ( { size, path } ) => {
                                        //      uk.gov.example - so split by "."
                                        const levels = path.split('.'),
                                        level = levels.length - 1,
                                        leaf = levels.pop(),

                                        //      By default, keep everything collapsed except the TLDs
                                        collapsed = (level > 0) ? false:false,
                                        id = path,
                                        childLinks = [],        //      All children empty to start
                                        parent = levels.join('.');

                                        //      Convert hover text from "/com.example.www" to "www.example.com"
                                        const name = path.split(".").reverse().toString().replace("\/","").replaceAll(",",".")

                                        //      Set up the node
                                        const node = {
                                                id,
                                                path,
                                                name,
                                                collapsed,
                                                childLinks,
                                                leaf,
                                                size: +size || 20,
                                                level
                                        };

                                        //      Add the individual node to the array
                                        nodes.push(node);

                                        //      Set up the links based on hierachy
                                        if (parent) {
                                                links.push({source: parent, target: path, targetNode: node});
                                        }
                                })

                                // link parent/children
                                const nodesById = Object.fromEntries(nodes.map(node => [node.id, node]));

                                //      Go through the links and add them to the childLinks
                                links.forEach(link => {
                                        nodesById[link.source].childLinks.push(link);
                                });

                                //      A tree with only specific nodes visible
                                const getPrunedTree = () => {
                                        const visibleNodes = [];
                                        const visibleLinks = [];

                                        (function traverseTree(node = nodesById[rootId]) {
                                                visibleNodes.push(node);
                                                if (node.collapsed) return;
                                                visibleLinks.push(...node.childLinks);
                                                node.childLinks
                                                        .map(link => ((typeof link.target) === 'object') ? link.target : nodesById[link.target]) // get child node
                                                        .forEach(traverseTree);
                                        })(); // IIFE

                                        return { nodes: visibleNodes, links: visibleLinks };
                                };

                                //      Set up the graph
                                const elem = document.getElementById('3d-graph');
                                const Graph = ForceGraphVR({"controlType":"orbit"})(elem)
                                        .graphData(getPrunedTree());

                                Graph.dagMode('td')                                     //      Top Down
                                .dagLevelDistance(50)   //      May be useful to dynamically adjust?
                                // .enableNodeDrag(false)        //      Don't let the nodes be draggable
                                .linkDirectionalParticles(1)
                                .linkDirectionalParticleWidth(2)
                                .linkWidth(1)
                                .linkOpacity(1)
                                .linkColor('red')
                                // .nodeAutoColorBy(name)
                                .nodeColor(node => !node.childLinks.length ? 'green' : node.collapsed ? 'red' : 'yellow')
                                .onNodeHover(node => elem.style.cursor = node && node.childLinks.length ? 'pointer' : null)
                                // .onNodeClick(node => {
                                //      //      Expand or Collapse the node
                                //      if (node.childLinks.length) {
                                //              node.collapsed = !node.collapsed; // toggle collapse state
                                //              Graph.graphData(getPrunedTree());
```

Screenshot of a code listing.

# Annex 3 - Questionnaire and Results

## 3.1 Preamble

The preamble is based on the in-house user-research guidance.

Thank you for volunteering to take part in my MSc research project today. I'm talking to a number of different people to understand their experiences of Virtual Reality and the Metaverse. This research typically takes less than 15 minutes.

I'll be asking you some questions and I would also like you to take a look at a prototype VR experience to see how well it works for you. I'm not testing you - there are no right or wrong answers - I'm interested in understanding how this idea performs. You can't break it - it's not a real service and it doesn't collect any data.

Some people report symptoms of dizziness, eye strain, or nausea when using VR. If experienced, these symptoms typically fade quickly. You can choose to stop at any time. I'm going to ask you to think aloud while you're completing the tasks. I'll be taking notes and will be recording what you see. Do you have any questions so far?

## 3.2 Questions

These questions were adapted from "A Technology Acceptance Model Survey of the Metaverse Prospects" ([Aburbeian, Owda and Owda, 2022](#)).

### 3.2.1 Pre-Experience Questions

Before we get started there are a few questions.

| What is your previous experience with Virtual Reality Systems? | | | | |
|---|---|---|---|---|
| I have never used it | I have rarely used it | I am an occasional user | I am a regular user | I am an experienced user |
| Comments: | | | | |
| What is your opinion on The Metaverse? | | | | |
| Very negative | Mostly negative | Neutral | Mostly positive | Very positive |
| Comments: | | | | |
| Others' opinion about the Metaverse affects my intention to use it. | | | | |
| Strongly disagree | Mostly disagree | Neutral | Mostly agree | Strongly agree |
| Comments: | | | | |
| I can't wait to try Metaverse | | | | |
| Strongly disagree | Mostly disagree | Neutral | Mostly agree | Strongly agree |
| Comments: | | | | |

### 3.2.2 Instructions

- There are two buttons you will use in this experience.
- The trigger controlled by your middle finger. This is called grip.
- The X button controlled by your thumb. This is called expand.

### 3.2.3 In-Experience Questions - Part 1

1. Look around. Describe what you can see.
2. Move your left hand around, what do you notice?
3. Squeeze your middle finger on the grip button. Holding it down, move your hand left. What do you notice?
4. Squeezing the middle button and moving your hand moves you through the experience. Move closer to the Node marked "UK".
5. Highlight the UK node and click it using your thumb's expand button. What do you notice?
6. Look around the tree. What do you see?

### 3.2.4 VRSQ

I'm now going to ask you a few questions about how you're feeling.

| Are you experiencing any of the following? | | |
|---|---|---|
| | Yes (describe) | No |
| General discomfort | | |
| Fatigue | | |
| Eyestrain | | |
| (Original) Difficulty focussing<br>(Asked) Difficulty concentrating | | |
| Headache | | |
| (Original) Fullness of head<br>(Asked) Pressure inside your head | | |
| Blurred vision | | |
| Dizzy | | |
| Vertigo | | |

### 3.2.5 In-Experience Questions - Part 2

1. Move to gov.uk, click expand, what do you notice?
2. Move to small-tc.gov.uk.
3. Move your hand so the white ball touches the small-tc node. What do you see?
4. Move to HMRC.GOV.UK. What was the experience like?
5. Expand HMRC.GOV.UK and look at each node. What do you see?
6. Expand each node in HMRC. What do you see?

### 3.2.6 Post-Experience Questions

Thank you. There are a few more questions. You may now remove the headset.

| Time passed quickly when using VR devices. | | | | |
|---|---|---|---|---|
| Strongly disagree | Mostly disagree | Neutral | Mostly agree | Strongly agree |
| Comments: | | | | |
| The Metaverse experience is exciting. | | | | |
| Strongly disagree | Mostly disagree | Neutral | Mostly agree | Strongly agree |
| Comments: | | | | |
| Using Metaverse will be helpful to the organisation. | | | | |
| Strongly disagree | Mostly disagree | Neutral | Mostly agree | Strongly agree |
| Comments: | | | | |
| I intend to use Metaverse in the future. | | | | |
| Strongly disagree | Mostly disagree | Neutral | Mostly agree | Strongly agree |
| Comments: | | | | |

3.3 Further information

- Do you have anything else you'd like to say about the experience?

3.4 Questionnaire Results

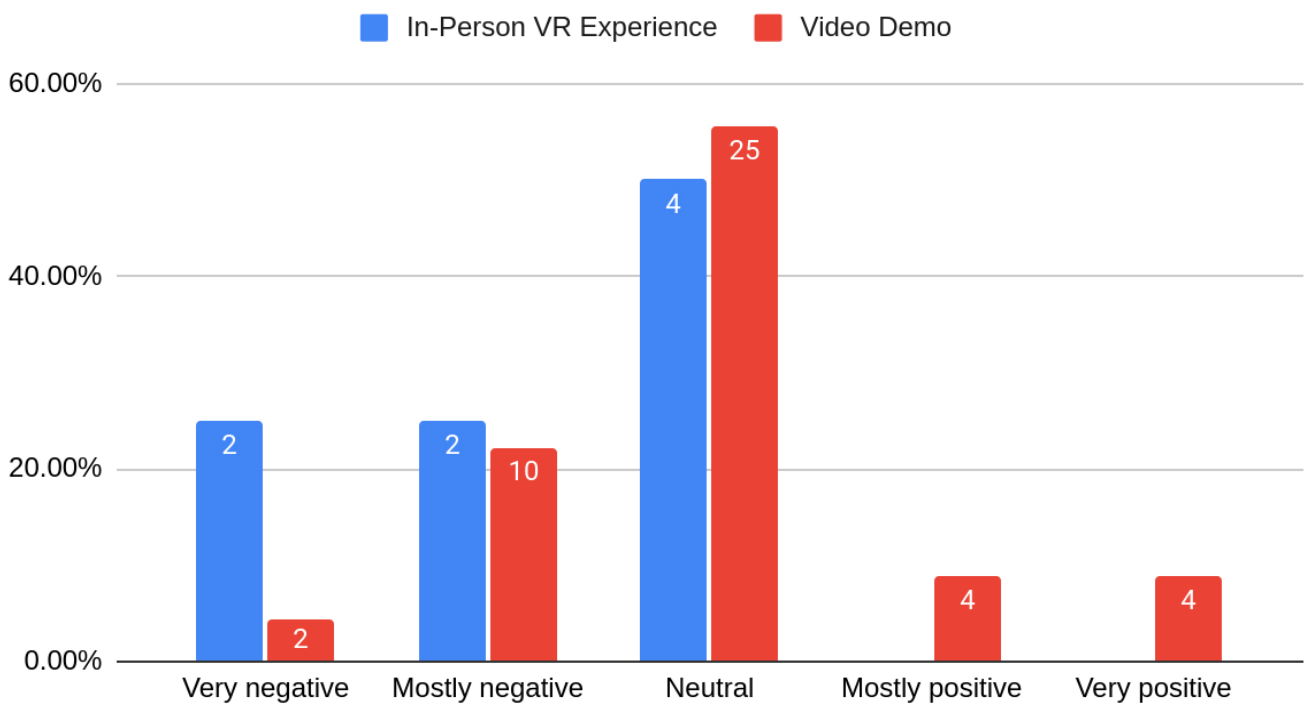

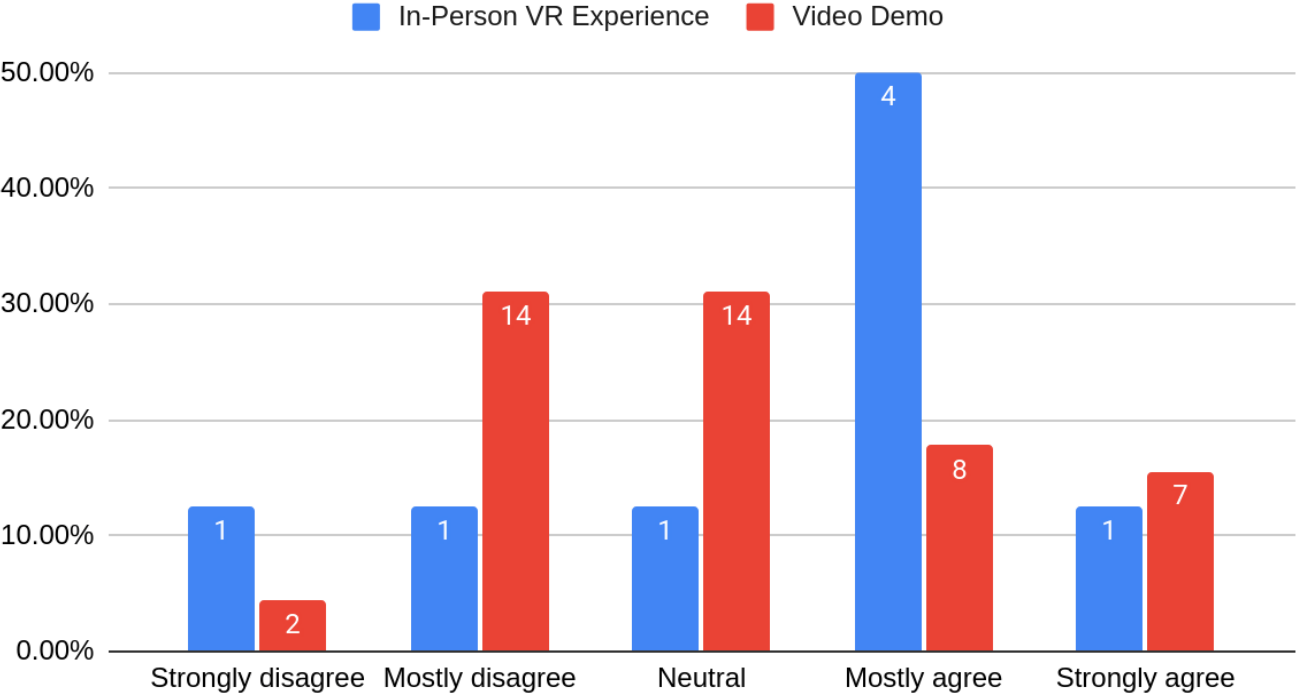

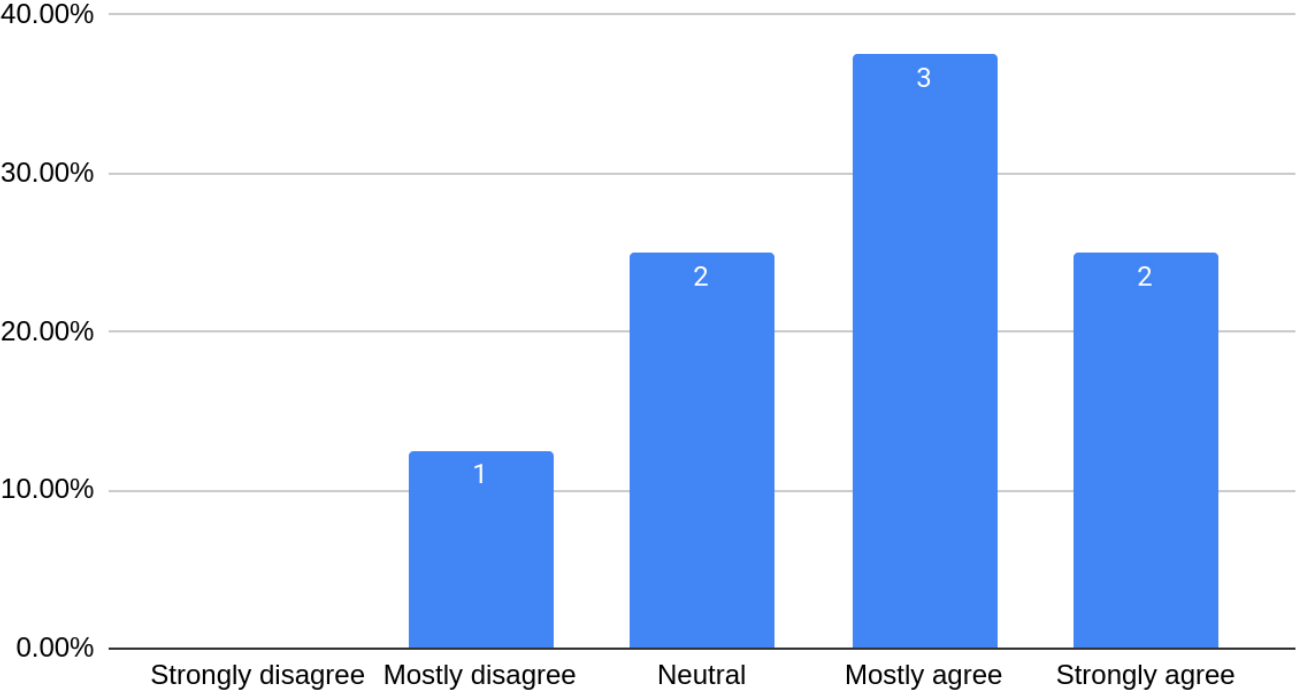

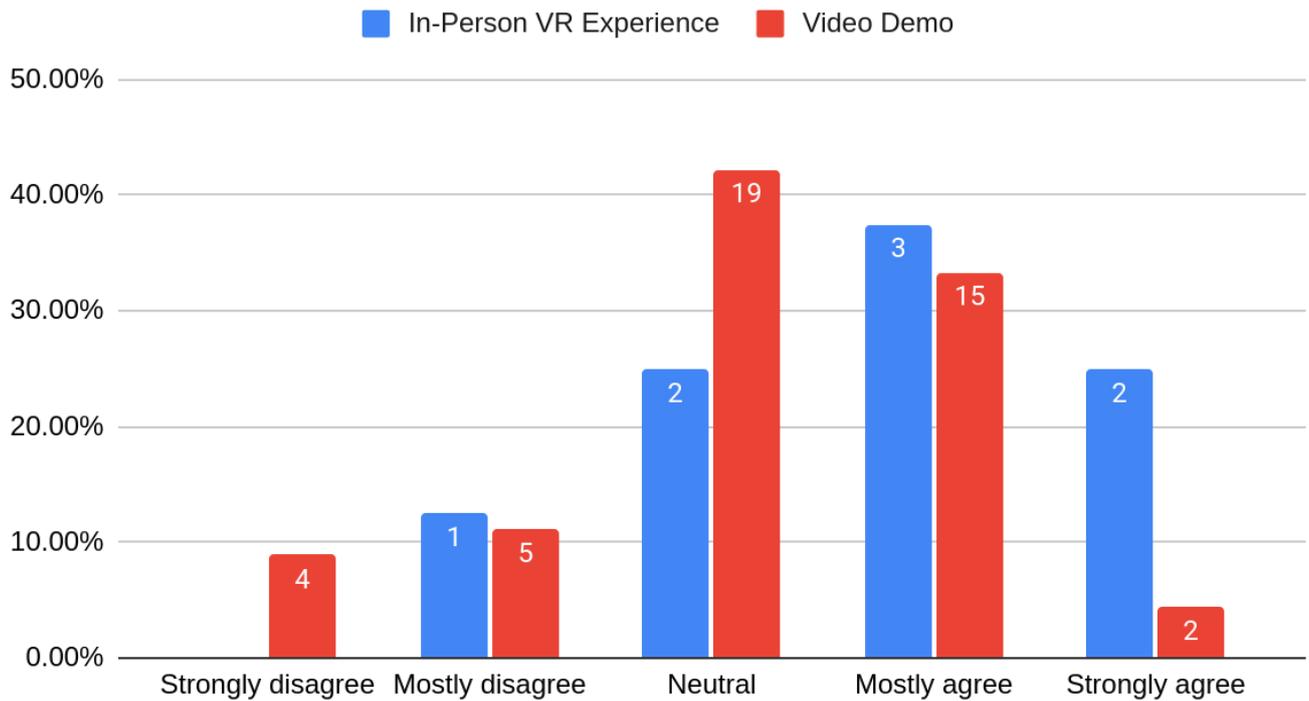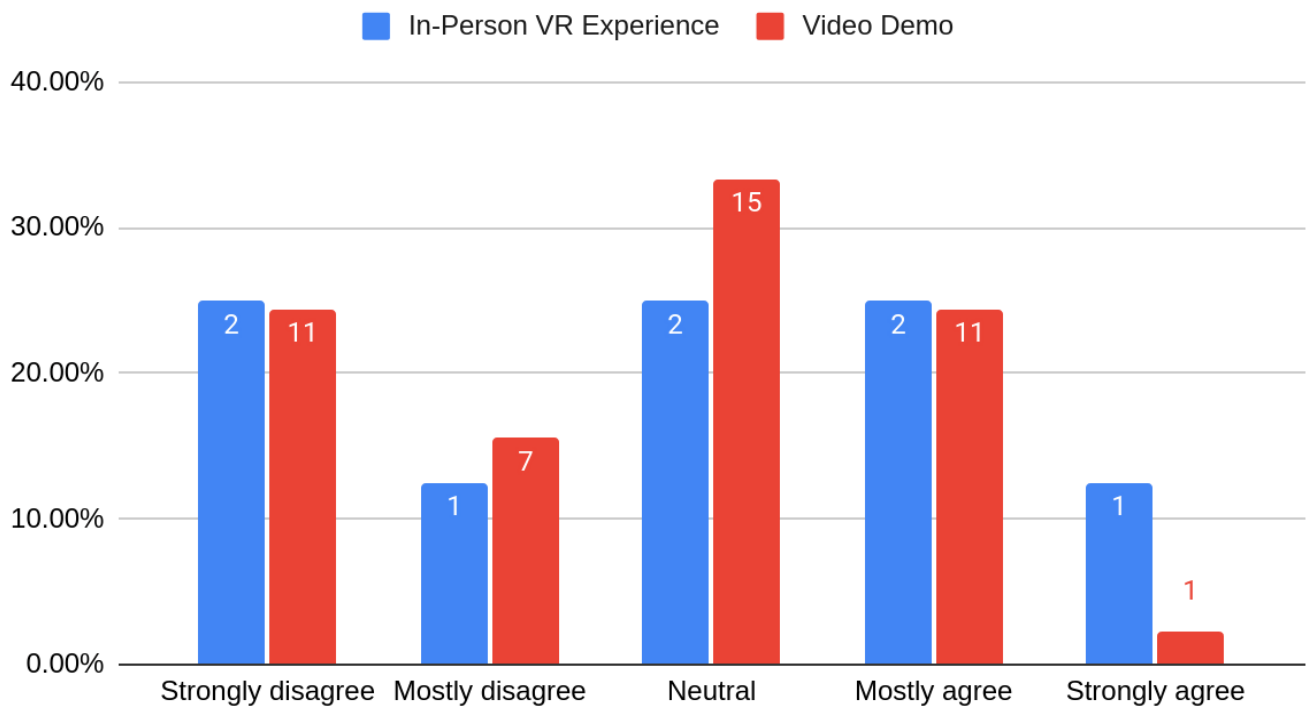

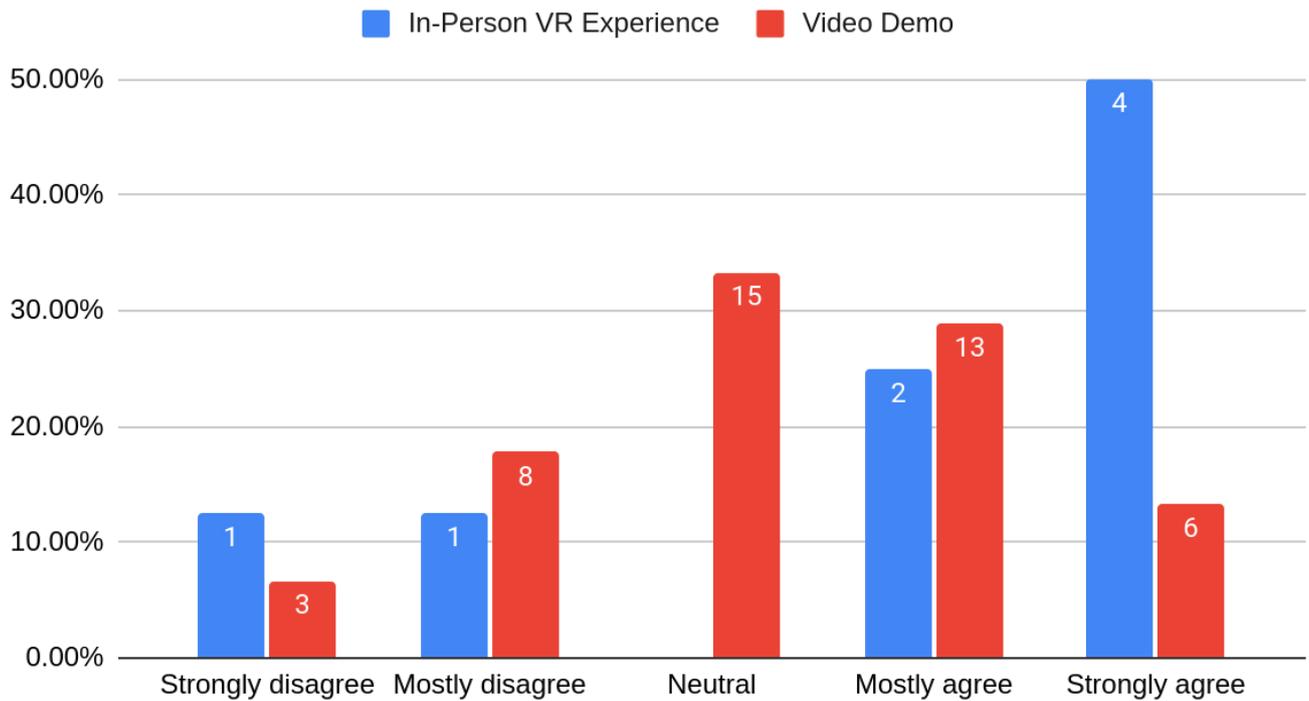

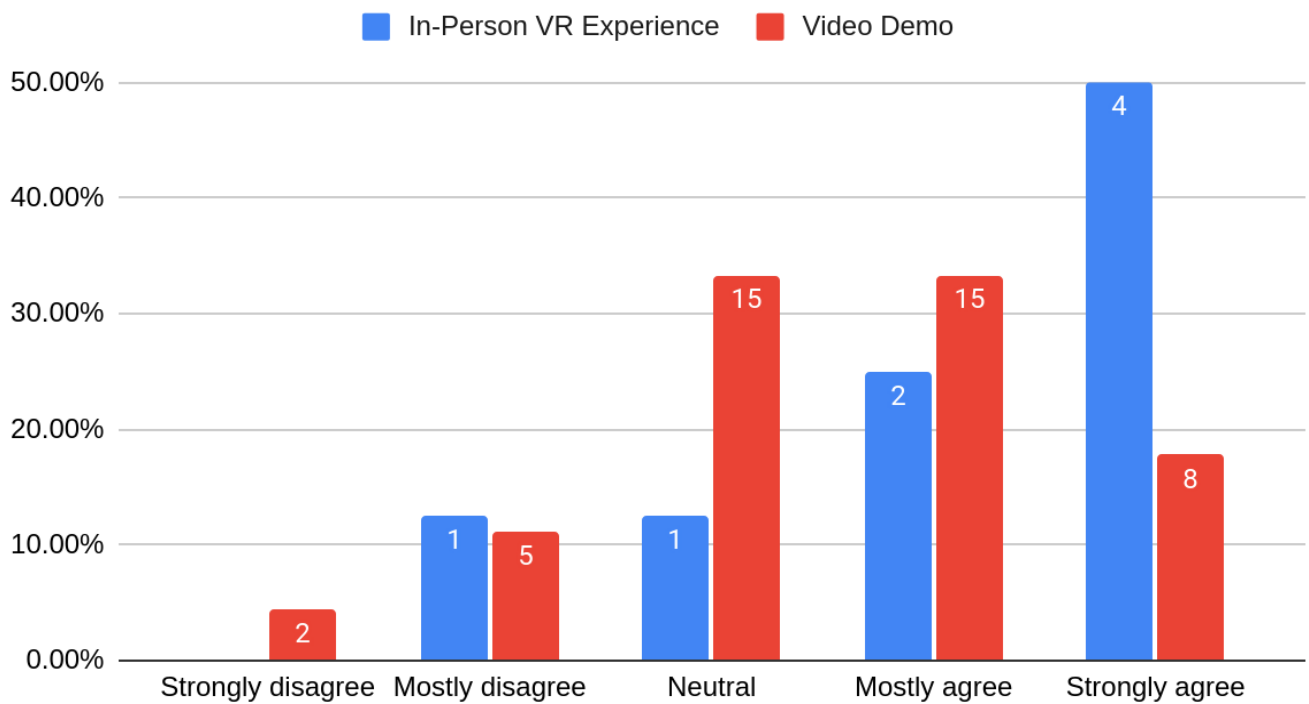

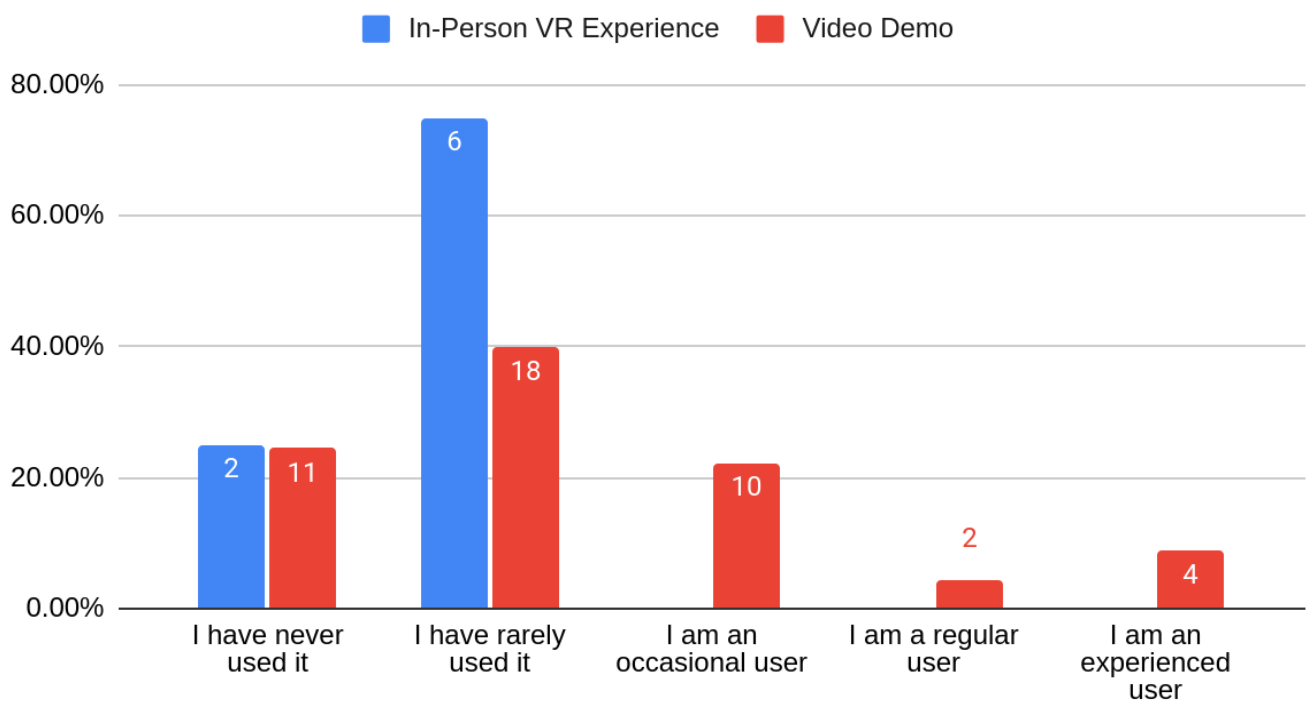

3.5 Sentiment Quantified Comments

Entries in italic are from participants from the group which tested the prototype.

| Question: What is your opinion on The Metaverse? | | |
| --- | --- | --- |
| Negative | Neutral | Positive |
| Scary, would prefer it didn't exist even if it has benefits | I don't know enough to have formed an opinion | Positive if there is an agreement on what it actually is. Seems to be being thrown around as the latest buzzword right now |
| I guess it's because it's becoming synonymous with Facebook, and… well… Facebook. [though to me it's William Gibson, Jaron Lanier, etc.] | neutral. seems revolutional | Going to be huge |
| Risk of centralised corporate application of VR. | context-dependent when overall +ve/-ve | open to be persuaded |
| I associate it strongly with Meta, who I do not trust. Also most screenshots/videos I've seen so far look like Runescape. | Looking to see where the hype lands | has a lot of potential |
| Entry barrier perhaps? | Little interaction but interested | Has potential, still not found a killer app yet though |
| I really don't like that Facebook are trying to own it. I don't like Facebook. | A more interactive internet where the best will be better and the worst will be even worse. | Nice I could play with that |
| Interested to see what's possible, but sceptical at the moment | It sounds like nothing else we've seen before, but I don't know much about it at this point | |
| Haven't tried it but everything I've seen about it has been negative. It seems like another addictive thing that will take you way from reality. I worry about the future where we're using it 24/7 | Pretty neutral. All I know about it really is from fiction. | |
| It's just stupid. There's no point to it. A waste of money and time. Trying to make something which isn't useful to people. Also trying to monopolise on the VR industry - it's not good. | I don't anything about it other than the hype | |
| I think that anything can be mapped onto 2D. And it's just another thing. Just because it exists that's no reason to do something with it. I think it will be exploited for more time wasting activities | Not formed an opinion | |
| I'm worried about security. Especially of young people. Possibly because I don't know | Open world thing. Can't put it into words. Get confused and think of the multiverse | |
| much. But the rules around security seem quite loose. | | |
| I can see the potential advantages. I am an avid sci-fi fan. So versions of the metaverse is something I'm very familiar with. I'm interested in how something commercially driven will end up | There are lots of benefits of the metaverse educationally. A lot of good things. Somewhere where people can get together where they couldn't in person. Music concerts. Escapism is very important. But the downside is that you can use it to escape real life. Not putting your focus on reality. You can prioritise your metaverse life over your normal life. Like Instagram where you're only uploading the best bits of you. | |

| Question: I can't wait to try The Metaverse | | |
|---|---|---|
| Negative | Neutral | Positive |
| There are no good headsets of a doable value. | Once I know more, I might be! | I think VR and AR are the future so anything that will bring this into the mainstream is an exciting development |
| Suspicious of data gathering activities (eg Meta) | Currently neutral, more concerned of side-effects in society | multiplayer games are going to be awesome |
| A stretch for me given I don't engage in social media | Until now I've not been hugely interested in the Metaverse (I spend a lot of time just on Instagram as it is!) but this may be due to my lack of understanding of it | Sounds interesting and a new way of seeing and understanding data |
| It would be interesting but I am sceptical how it will work, how much it will cost. Will the hardware be available to support it. | I'm curious. I want to take a look. | Always exciting for something new. |
| Don't have any burning desire to try it or buy a headset. IT feels like 3D tv in the home. It was a thing until it wasn't a thing. It feels like this could be the same. It might not be. | I've got mixed feelings. I don't think it is a necessity. We have created something that may prove to be good but will take years. I'm a busy person so I'll leave it to other people | It is exciting and a new possibility |
| I don't care about the metaverse. It's just not worth my time. Whatever the metaverse can do a normal computer application could do. It sounds boring. It's a gimmick | | Looks an interesting new tool |
| I'm open to new experiences. If something came up - a use case - I'd crack on and use it. I'm slightly concerned that it will be exploited for entertainment rather than utility. | | Want to see what's going to be there. Always interested in getting one. But never got around to it. Have friends who play and then they stop. Good at the start. Virtual parties and talking to people. Virtual concerts. Looking around. That appeals to me. Gaming - I play here and there. |
| | | I love technology. I just feel like you're somewhere else when you're not. It's a different way to learn. I'm a doer not a reader - being able to see it with my own eyes will be amazing. I'm just really excited. |
| | | Excited to see how it works and what it can do. |
| | | I'm a technology geek. I'm genuinely interested in the application of new technology. And science fiction! |

| Question: The Metaverse experience is exciting. | | |
| --- | --- | --- |
| Negative | Neutral | Positive |
| Devalues what it means to be human, too many risks as humans are inherently evil | It could well be… | Great to see some VR applications for data! |
| Not sure it adds to understanding | It was interesting, but I wouldn't say exciting. Useful tool for work purposes. A different way to see things. | interesting research |
| it doesn't feel much different to previous attempts at 3d visualisations | I don't know. Why Metaverse and not VR. Metaverse is something very specific created by other people. | It looks cool, and I can certainly see how it expands to make it something you can't get elsewhere |
| I think it's just not so much fun as it is made out of it. But you could do some cool things. But you could replicate it on a normal screen with normal controls | Lots of use of opportunities but access to hardware may be limiting. Ability to visualise things is very powerful | Opens up a new way to interact and share data |
| | | I wasn't aware of these potential use cases, or that organisations could load in their data or connect it to the Metaverse (I've not been paying nearly enough attention!) |
| | | Level of interaction you can't get anywhere else. Immersive. Easy to explore different sections. Unique experience. Can't do that anywhere else. |
| | | A different way of seeing things. Instead of seeing things in a text format - I could see visually and go into depth with the things that you wanted. Very visual. |
| | | It was all new. I was enjoying having a look round and seeing where everything was. I got lost in it all. |
| | | I can see it being a good tool to visualise complex issues to senior people. |
| | | Because the tech has reached the point where it is sufficiently seamless. Imperceptible lag. It feels natural to operate in. Our animalistic brains perceive it well. Much more natural dealing with lists. |

| Question: Using The Metaverse will be helpful to the organisation. | | |
|---|---|---|
| Negative | Neutral | Positive |
| This could be done on a flat screen. I don;t know how easy it would be to manipulate the data (but I am sure this will improve in time) | It might be… | I think right now, it's early days but we should be starting to think about and planning for future applications |
| large cash outlay, see above re not different enough | Probably not for our kind of data but can see value in others. | Possibly for some things 3d is useful where there are cross links |
| At the moment the usage seems descriptive rather than interactive - anything interactive I've seen (eg Zuckerberg demos) is far short of anything useful | It depends | Create a shared space with an easier environment for people to visualise as a non-expert than excel and spreadsheets |
| While I have no doubt that there are strong use-cases for VR in business, my concern is that this could turn into a case of Doing Metaverse Because Everyone Is Doing Metaverse, and ultimately trying to shoehorn it into applications where it isn't the most appropriate tool. | not sure how much it adds compared to 2D, although it might work for e.g. people survey. Or using it for 3D charts (not hierarchies). | Visualisation can improve confidence |
| Depends on the purpose . I don't see govt employees hopping into the Metaverse and faffing about. I do think that something like this could be on a normal computer. | I think it could be but it's hard to say what for at the moment. The domain management thing, like seeing which has issues to investigate, I still think it's more efficient to have something like a 2d report that sends an email alert or something saying there's something wrong with a domain | Our dataset is complex. Analysts have trouble navigating it. Using the metaverse would help them I think and give senior people an appreciation of it's complexity. |
| | I can think of an example where we store certain information as a hierarchy, so your specific example actually might be useful. But we've done fine without it before. | I can't think of a use case today but the principle of being able to map and model data like this is phenomenal. It needs developing but as a basic concept turning highly complex data into something a human can interact with is amazing. Imagine visualising a family grouping and wider contacts of a child in social care or between families. Are there individuals deep within a family network connecting to other children etc. |
| | Unsure. Different people have different requirements. | My organisation has some big (as in millions) hierarchical datasets with very uneven content. It would interesting to apply this approach to some of those |
| | There is potential | Provides new ways to illustrate structures/problems that the organisation is facing. Can already imagine how it will make it easier to compare how certain domains/tables/ |

|  |  | datasets are more complicated or bigger than others. It may make it easier for users to find issues or insights. I think it's easier to pull insight from a top level, though I question how helpful it will be at a granular level. |
|---|---|---|
|  | Such a short experience. Looked like it had potential. But how would it work in practice? And the difficulty of scaling it and using it for long periods of time. I can't see how it would make something more efficient. I can see how it is better to present something or explain something. But in terms of trying to fix something I can't see it. More like a presentational tool. | More an more data is flooding into the organisation. I think this will help us to explore the data in new ways. |
|  | At this point, I'm not sure. I only saw the surface of it. I don't know what that was for. It was showing me information, but how I would use it properly I'm not sure. | Depends on how you design the system or product. There are areas where you can create and learn. Can break down the tree and see exactly where it is. Good for demoing stuff. Better than a conversation. |
|  | The question is too high level to give a definitive answer. For analytic and design. Less so for economics. I might visualise strategy. It can help us tell stories at a pace and way that people will find engaging. It is very tempting. I've always wanted to get inside the data. Being able to visualise the sharing agreements it just plays beautifully. For people who deal with lots of text, less so. | Based off that experience. When explaining DNS and how a tree works and subdomains sit - it can be quite difficult. I'm a visual learner. Being able to jump into a diagram made it easier to understand and more interesting. Things which are more interesting are better for earning. The highlighting really helped to focus my attention and go into depth |
|  |  | So different viewing something in that way. I'm used to just a laptop. If I was doing this, I'd be more interested in how I can utilise it. |
|  |  | Because having this virtual representation of things helps with processing information. Makes it easier. Especially colours make you see something is going on. Better than spreadsheets and static information. |

| Question: I intend to use The Metaverse in the future. | | |
|---|---|---|
| Negative | Neutral | Positive |
| I'm a coder, so don't tend to consume data too much. | Can't see applicability to me at the moment, but I'll keep an open mind | I initially didn't expect this is what the meta verse could look like. I assumed it would be more linked to VR type games with characters. Excited to see how we could use it in the future |
| Probably not enough value. | It's a novelty, I'll try it if someone has the equipment and has already set it up for me. | possibly but only using as a tool in specific use cases |
| it would be cool to use it, but I doubt it would be practical or possible to implement within the civil service (given that it's taken 2 years to get access to R packages!) | only in specific scenarios | I do not know if I will have the opportunity, but would like to. |
| I find 3D visualisations make me nauseous | I am waiting for the metaverse to get bigger or the price to come down. | Because it is happening and I am interested |
| I'll take the opportunity if I get it. That doesn't seem likely in a work context. | | I can see how this would be useful in visualising the connections between objects/data in our infrastructure |
| All I think about is Facebook. When I hear metaverse all I hear is an extension of facebook. And I have no interest in that. If you mean things separate to that then yes, potentially. LD7111<br><br>No one says search, everyone says Google - it feels like that's happened with Metaverse. It is worrying. They are looking to get people addicted and monetised. IT's not coming from a place where they want to help humanity - it is just business expansion. Especially when this could be quite addictive. | | think it will be useful for showing data visually in 3D, at least will help get an understanding of scale |
| It's a gimmick. I don't see the point. | | Could be useful for Linked data. |
| | | I've already been building and playing with Metaverse apps. |
| | | I don't have the technical skills now to develop this but this is very exciting! |
| | | Lots of potential for collaborative work, spatial placing etc. |
| | | It would be beneficial for a range of people to understand the scale and complexity of the data that my organisation works with |
| | | Always been keen. Now, on seeing this, it has motivated me to look into it. |

| | | | |
|---|---|---|---|
| | | | It interests me. It is new so it is something that is exciting for me and I want to know more about it and how it would benefit me. |
| | | | It is a fun way of exploring complex ideas |
| | | | I've seen it now. This is a good use. Metaverse is huge and has many different uses and scope. For me and my job I can see the benefit of something about. |
| | | | It's just natural. It's part of evolution to act in virtual spaces. |

| Question: Any further thoughts on what you've seen? | | |
|---|---|---|
| Negative | Neutral | Positive |
| Not sure what the interactions between people adds in this case | It worked well. I felt like you really had to reach. Would be better if it worked remotely. While it wasn't a problem. You get very deep into something quickly, so you lose the big pyramid quickly and lose track. | Very cool |
| Does Facebook get my data? | I don't think we exploited the dynamics enough. I didn't experience the all the possibilities | It's interesting. Graphics needs development |
| I'm reluctant because of what I've heard. Environments you go in and there are no rules and people come and attack you. Kids' safety is very important. Adults too. | | Seems so exciting but no idea about how I would even start to look at ways to involve it in my work, other than waiting for it to be main system. |
| Having removed the headset – I felt it was a bit heavy. | | I would like to do some work in this area myself with geospatial data representation |
| | | Interesting - would like to discuss more, we are actively looking at how to use metaverse in policy making |
| | | Should be particular use cases where this way of immersive UX makes workflows more efficient. Does this has an efficient visual search engine? |
| | | Really clear explanation, and I love the interactive demo; really brings the idea to life in a way that a static slide could never do. |
| | | Different people have different data needs, so would need different visualisations. The more we can empower people to create their own visualisations, the more powerful this tool will be |
| | | Pretty unique. Not come across anything like that. It was fully immersive. Definitely something I want to try again. An awesome piece of tech. |
| | | It was really cool. It was fun. I've never tried it before. I didn't feel silly using it. I enjoyed using it. I want to know more! I've only known it as a gaming thing because of my other half. But to potentially use it for work is really interesting. |
| | | You could have chosen to put that in a neutral featureless space. Your attention would focus on the construct. Having a room and things to focus in the distance helped the user experience. I had reference points to prevent motion sickness. I thought it would be distracting and it wasn't. It helps to ground |

| | | you in the artificial construct. I was surprised how intuitive it became. |
|---|---|---|

## 3.6 Individual Informed Consent Form

**Informed Consent Form**

| Title of Research | EXPLORING VISUALISATION OF CRITICAL CYBERSECURITY DATA USING THE METAVERSE |
|---|---|
| Name Researcher | TERENCE EDEN |
| Name of supervising academic (where appropriate) | SHARON SIBANDA |
| Address for correspondence | CDDO<br>WHITECHAPEL BUILDING<br>LONDON<br>E1 8QS |
| Telephone | |
| E-mail | |
| Description of the broad nature of the research | To discover whether a range of stakeholders will find using the Metaverse an acceptable proposition for a modern organisation, how much training they need to navigate complex data, and whether it will influence them to conceptualise new ways to interact with other business systems. |
| Description of the involvement expected of participants including the broad nature of questions to be answered or events to be observed or activities to be undertaken, and the expected time commitment | Participants will be interviewed to assess their familiarity with Virtual Reality environments.<br>They will be asked to spend some time evaluating data within a Metaverse construct.<br>A qualitative survey and further interview will then be carried out.<br>Although there are no sensitive topics, participants will be informed about the potential medical risks of using VR (primarily motion sickness, eye strain, and epileptic reaction). All participants will be free to withdraw at any time. |
| Additional information about the research | Part of a Cabinet Office funded MSc |

Information obtained in this study be anonymous (i.e. individuals and organisations will not be identified *unless this is expressly excluded in the details given above*).

Data obtained through this research may be reproduced and published in a variety of forms and for a variety of audiences related to the broad nature of the research detailed above. It will not be used for purposes other than those outlined above without your permission. Participation is entirely voluntary and participants may withdraw at any time.

Northumbria University is the data controller under the Data Protection Act (1998)

*By signing this consent form, you are indicating that you fully understand the above information and agree to participate in this study on the basis of the above information.*

**Participant's signature**                                                                 **Date**

*Please keep one copy of this form for your own records*

Form asking people to consent to being interviewed and experimented upon.

# Annex 4 - Presentation to DataConnect22

Video of the presentation:

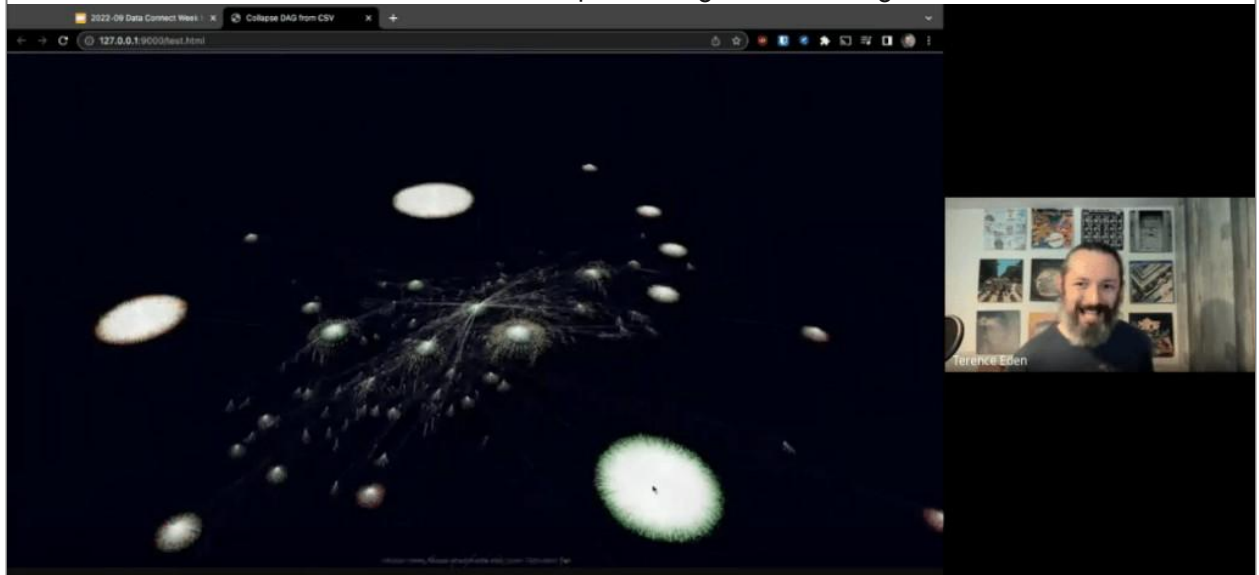
Screenshot of Terence presenting via streaming video.
https://www.youtube.com/watch?v=TEx_JrZagAM

# Copyright and Copyleft

This document is Ⓒ Terence Eden [Creative Commons — Attribution-NonCommercial 4.0 International — CC BY-NC 4.0](#).

It may not be used or retained in electronic systems for the detection of plagiarism. No part of it may be used for commercial purposes without prior permission.

Any source code is under the [MIT Licence](#).

This document contains public sector information licensed under the [Open Government Licence for public sector information v3.0](#).

---

1. This has the effect of formatting DNS into the order now favoured by Sir Tim Berners-Lee, the inventor of the World-Wide Web ([Runciman, 2011](#)).↩